\ifdefined\XeTeXrevision\else\pdfoutput=1\fi
\documentclass[11pt,a4paper]{article}

\usepackage[T1]{fontenc}
\usepackage{lmodern}
\usepackage[margin=2.5cm]{geometry}
\usepackage{amsmath,amssymb}
\usepackage{graphicx}
\usepackage{booktabs}
\usepackage{array}
\usepackage{xcolor}
\usepackage{cite}
\usepackage{placeins}
\usepackage[colorlinks=true,linkcolor=blue!50!black,citecolor=blue!50!black,urlcolor=blue!50!black,pdftitle={Symmetry breaking from conformality loss on holographic R\'enyi defects},pdfauthor={Siqing Yu}]{hyperref}

\numberwithin{equation}{section}
\newcommand{\Tc}{T_c}
\newcommand{\rhob}{\bar\rho}
\newcommand{\Bb}{\bar B}

\newcommand{\ee}{\mathcal{E}}
\newcommand{\Om}{\Omega}
\newcommand{\SF}{{\rm SF}}
\newcommand{\N}{{\rm N}}

\newcommand{\Nn}{\mathcal{N}}
\newcommand{\muc}{\mu_c}

\newcommand{\Dz}{\mathcal{D}_0}
\newcommand{\Dsb}{\mathcal{D}_{\rm SB}}
\newcommand{\Dh}{\hat\Delta}
\newcommand{\tilt}{\hat t}
\newcommand{\uh}{u_h}
\newcommand{\ve}{\varepsilon}

\begin{document}

\hypersetup{pageanchor=false}
\begin{titlepage}
\begin{center}
\vspace*{1.5cm}
{\LARGE\bfseries Symmetry breaking from conformality loss\\[0.2em] on holographic R\'enyi defects\par}
\vspace{1.2cm}
\renewcommand{\thefootnote}{\fnsymbol{footnote}}
{\large Siqing Yu\footnote{\href{mailto:yu@siqing.ai}{\texttt{yu@siqing.ai}}}\par}
\vspace{0.3cm}
{San Diego, California, USA\par}
\end{center}
\vspace{1.0cm}
\begin{abstract}
\noindent
The R\'enyi entropies of a ball are computed by the R\'enyi defect (the twist operator viewed as a conformal defect), which in holography is dual to a hyperbolic black hole. We identify hyperbolic black holes with homogeneous charged scalar hair as conformal defects that break the U(1) symmetry. At large $N$ and for repulsive self-interactions, conformality loss of the R\'enyi defect turns explicit symmetry breaking into spontaneous breaking: below a critical index $n_*$ the symmetry-breaking defect is reached only with an explicit source on the entangling surface; at $n_*$ the dimension of a defect operator from a light charged scalar turns complex; beyond $n_*$ the R\'enyi defect flows to the symmetry-breaking defect with no source. The condensate, expected to break the symmetries of hyperbolic space, instead preserves them outside a layer at the entangling surface, whose width diverges at $n_*$ with Berezinskii--Kosterlitz--Thouless scaling. The universal term of the R\'enyi entropy therefore jumps at $n_*$ to that of the symmetry-breaking defect. We demonstrate this mechanism in the probe Abelian Higgs model in AdS$_5$ at the Bogomolny point, with closed-form saddles and defect data, and numerical flows. For a ball we prove that a positive source-free flow exists exactly when $n>n_*$ and the radius is large enough, and that it ends at the symmetry-breaking defect as the radius grows.

\end{abstract}
\end{titlepage}
\hypersetup{pageanchor=true}
\setcounter{footnote}{0}

{\setlength{\parskip}{0.2em}\tableofcontents}
\newpage

\section{Introduction}
\label{sec:intro}

The R\'enyi entropies $S_n=\frac{1}{1-n}\ln{\rm Tr}\,\rho_A^n$ of a region $A$ are computed by a twist operator, a codimension-two defect on the boundary $\Sigma$ of $A$ around which the $n$ replicas are permuted. In a conformal field theory (CFT) the twist operator of a planar or spherical $\Sigma$ is a conformal defect \cite{HMS2014,BMMS2015}; see \cite{BGLMM2016} for defects in CFT. It has its own spectrum of defect operators and its own defect data, such as the conformal weight \cite{HMS2014,Perlmutter2013}, the normalization of the displacement operator and, for a surface in four dimensions, the coefficients of its Weyl anomaly \cite{HMS2014,BMMS2015,BDF2016,JensenOBannon2015}. Closely related are monodromy defects, around which the fields return up to a symmetry transformation \cite{GMP2013,BCPRS2021,GHJK2021}. For a ball of radius $R$ or a half-space, ${\rm Tr}\,\rho_A^n$ is mapped to the thermal partition function on $S^1\times H^{d-1}$ at the temperature $T_0/n$, with $T_0=1/2\pi R$ \cite{CHM2011,HMSY2011}. In holographic CFTs \cite{Maldacena1997,GKP1998,Witten1998} this partition function is computed by a hyperbolic black hole \cite{CHM2011,HMSY2011}, also with a chemical potential for a U(1) charge \cite{BHMMMS2013}. For general regions, the dual of the twist operator is a cosmic brane \cite{Dong2016}.

A R\'enyi defect, as we call the twist operator, need not preserve the global symmetries of the CFT. Below $T_0$, hyperbolic black holes can carry scalar hair \cite{DMRS2010,BMM2013}, and Belin, Maloney and Matsuura distinguished two instabilities. The constant mode on $H^{d-1}$ sets in at $n=1$ for a scalar of a particular mass. Its hair does not fall off toward the boundary of $H^{d-1}$, which is the entangling surface, and it computes R\'enyi entropies when operators inserted at the entangling surface source it. The normalizable mode sets in at a larger index $n_*$, and with a chemical potential its onset was located \cite{BHMM2014,Ren2020}. Its condensation was expected to break the symmetries of hyperbolic space \cite{BMM2013}, and Belin, Hung, Maloney and Matsuura described the phase beyond the onset as an impurity operator with a nonzero expectation value on the entangling surface \cite{BHMM2014}.

Neutral hairy hyperbolic black holes with backreaction are known analytically \cite{Ren2019} and numerically \cite{FHL2017}, and their R\'enyi entropies were computed there and in \cite{Bai2022}. Below $n_*$, such hair needs a source at the entangling surface. For the black holes in AdS$_4$, an extended R\'enyi entropy and the conformal weight were computed in \cite{Sun2023}. In field theory, the fixed points of a coupling on the conical singularity of the O($N$) model merge and disappear beyond a critical index \cite{MFS2009}. The R\'enyi defect of the large-$N$ O($N$) model in three dimensions becomes unstable toward symmetry breaking at $n=1$ \cite{HLS2018}, and lattice simulations at O(3) critical points see R\'enyi defects order at large $n$ \cite{ZWCY2026}. Two questions are therefore natural: what the hairy saddles are as defects of the CFT, and where the R\'enyi defect goes once the normalizable instability has set in. We show that the hairy saddles are conformal defects that break the U(1), and that once the R\'enyi defect loses conformality it flows to one of them with no source.

\paragraph{The mechanism.} In the conformal frame $S^1\times H^3$, in which $H^3$ plays the role of AdS$_3$, a Kaluza--Klein mode on the hyperbolic black hole is a defect primary, and a condensate that is constant on $H^3$ is a one-point function $\langle O\rangle=a_O/r^\Delta$ at the distance $r$ from the entangling surface \cite{GHJK2021,CuomoZhang2023}. A homogeneous hairy saddle is therefore a symmetry-breaking conformal defect, which we call $\Dsb$, distinct from the R\'enyi defect $\Dz$ of the normal phase, and both questions become questions about defect flows. Below $n_*$, $\Dz$ has a relevant charged operator, and $\Dsb$ is the infrared fixed point of $\Dz$ deformed by it, a pinning field on the entangling surface; this is the source invoked in \cite{BMM2013}. At $n_*$ the dimension $\Dh$ of this operator meets that of its alternate-quantization partner at $\Dh=1$, the defect analog of the Breitenlohner--Freedman (BF) bound \cite{BF1982a,BF1982b}, and beyond $n_*$ both are complex, $\Dh=1\pm i\nu$. This is a merger of defect fixed points followed by conformality loss \cite{KLSS2009}, the holographic counterpart of the merger in the O($N$) model, as for Wilson lines \cite{ACKMR2022,ACKMR2023}.

Beyond $n_*$, then, $\Dz$ is not a consistent defect, and the R\'enyi defect flows with no source to $\Dsb$. Conformality loss thus turns the symmetry breaking that a pinning field imposes below $n_*$ into spontaneous breaking beyond it. Both statements hold at large $\Nn$, with $\Nn$ proportional to the two-point coefficient of the current. The conformal group of the defect acts on $H^3$ by isometries, so a conformal endpoint of the flow is homogeneous on $H^3$, and the inhomogeneity is the flow itself. The phase expected to break the symmetries of hyperbolic space beyond $n_*$ \cite{BMM2013,BHMM2014} is the homogeneous saddle outside a boundary layer at the entangling surface, whose width $z_*$ is set by the cutoff $\delta$ through a scale of Berezinskii--Kosterlitz--Thouless (BKT) type, $z_*/\delta\propto\exp(\pi/\nu)$. As $\delta\to0$ the layer shrinks onto the entangling surface, so the coefficient of $\ln(R/\delta)$ in the R\'enyi entropy of a sphere is that of $\Dsb$, and it jumps at $n_*$.

\paragraph{The model.} We demonstrate the mechanism in the model of \cite{MainPaper}, the probe-limit Abelian Higgs model in AdS$_5$ \cite{Gubser2008,HHH2008a,HHH2008b,Herzog2010} with the scalar mass at the BF bound and the quartic coupling at the Bogomolny value. Its condensed phase is known in closed form on every static, uncharged AdS$_5$ background with maximally symmetric spatial sections, and on the hyperbolic black holes at $T=T_0/n$ it gives the homogeneous hairy phase, which at zero chemical potential exists for every $n>1$. At $n=1$ the charged operator is marginal precisely when $\Delta=d-2$, which in $d=4$ is the BF bound, so near $n=1$ it is weakly relevant. The defect data of $\Dsb$ are in closed form: its one-point coefficient, the change of its anomaly coefficient (negative, as the defect $b$-theorem \cite{JensenOBannon2015} requires), the change of its conformal weight and the normalization of its tilt operator. The merger is at the onset $n_*\approx6.58$ of \cite{BMM2013}. We prove that on a sphere no positive source-free solution exists for $n\le n_*$, and that for every $n>n_*$ exactly one exists once the sphere is large enough; it minimizes the action and, as the sphere grows, ends at $\Dsb$ in the interior and gives the universal R\'enyi term of $\Dsb$. The numerics add the layer, which the proof does not locate. They find the source-free flows for $6.59\le n\le30$, with the coefficient of $1/\nu$ in the logarithmic width of the layer equal to $\pi$.

\paragraph{Genericity.} The mechanism is not tied to the closed form. For a probe scalar at zero chemical potential the merger exists for $1<\Delta<3$, the window of \cite{BMM2013}. An argument, rigorous in the model and confirmed for three further values of $\Delta$, says that the source-free flow ends at a homogeneous saddle when the action of the probe scalar is bounded below. Since $n_*\to1$ as $\Delta\to1$, even the lowest R\'enyi entropies can lie beyond the merger: the second does for $\Delta<1.438$, and the third for $\Delta<1.658$. By this argument, their universal terms are then those of the homogeneous saddle. The transition at $n_*$ also requires that no source-free flow exist at $n_*$, which repulsive self-interactions guarantee. With an attractive quartic coupling stabilized by a sextic one, symmetry breaking can come before conformality loss: the R\'enyi defect then orders below $n_*$, at a scale that diverges as a power rather than exponentially.

\paragraph{Chemical potential.} The merger is not tied to zero chemical potential. At imaginary chemical potential the closed form gives conical defects with a U(1) flux and a winding condensate, with the structure of the monodromy-pinning defects of \cite{KravchukRadcliffe2025}. The merger persists there, on a line that rises from $n_*$ to $6.70$ at half a flux quantum, and beyond it, at the two fluxes computed, the R\'enyi defect with flux flows to the winding saddle. At real chemical potential the line continues to lower indices: the merger sits at the onset $\mu_{\rm norm}(n)$ of the normalizable instability \cite{MainPaper}, which reaches $n=1$ at $\mu_{\rm norm}R=1.39$. No source-free flow exists below the line, and a Lyapunov function shows that bounded positive flows beyond it approach $\Dsb$; near the line their walls follow the $\pi/\nu$ law.

\paragraph{Further results.} Classically in the bulk, the onset $n_*$ does not move at any order in the backreaction, and at first order the correction to the R\'enyi entropies of the hairy saddle is in closed form. At finite $\Nn$ the U(1) cannot remain spontaneously broken on a surface defect \cite{CuomoZhang2023}: with the closed-form tilt normalization, the order parameter decays along the defect as a power of the logarithm of the distance, with an exponent of order $1/\Nn$. A real scalar with a $\mathbb{Z}_2$ symmetry gives the same flows, and the theorem of \cite{CuomoZhang2023}, which concerns continuous symmetries, does not apply to it.

Apart from the backreaction and finite $\Nn$, all results are at leading order in $1/\Nn$ and in the probe limit. The saddles, their defect data and the first-order free energy are in closed form. The spectrum, including $n_*$, is numerical, except for its values and slopes at $n=1$ and its limits as $n\to\infty$. The flows are numerical too, except for the properties we prove on the sphere and at real chemical potential. The genericity results and the merger line with flux are numerical or rest on arguments.

Section~\ref{sec:renyidefect} recalls the model, sets up the defect dictionary and computes the spectrum of $\Dz$. Section~\ref{sec:defect} gives the defect data of $\Dsb$, and section~\ref{sec:flow} the flows to it and the universal R\'enyi term. Section~\ref{sec:generic} varies the quartic coupling, the scalar mass and the shape of the potential, and section~\ref{sec:backreaction} turns to chemical potential, backreaction and finite $\Nn$. Section~\ref{sec:discussion} relates the results to earlier work, states their scope and lists open questions. The appendices give the proof on the sphere, the backreaction at first order and the numerical methods.

\section{The R\'enyi defect and conformality loss}
\label{sec:renyidefect}

\subsection{The model and the defect dictionary}
\label{sec:setup}

We use the model, the conventions and the notation of \cite{MainPaper}: the Abelian Higgs model in AdS$_5$ in the probe limit, with the matter action
\begin{equation*}
S_{\rm matter}=\int d^5x\,\sqrt{-g}\,\Big[-\frac{1}{4e^2}F^{ab}F_{ab}-|D_a\Psi|^2-m^2|\Psi|^2-\lambda|\Psi|^4\Big],\qquad D_a=\nabla_a-iA_a,
\end{equation*}
the scalar mass at the BF bound, $m^2L^2=-4$, and the quartic coupling at the Bogomolny value $\lambda=e^2/2$ unless stated otherwise. The background is the hyperbolic black hole,
\begin{equation}
ds^2=\frac{L^2}{u^2}\Big(\frac{du^2}{f(u)}-f(u)\,dt^2+R^2dH_3^2\Big),\qquad f(u)=1-\frac{u^2}{R^2}-\Big(1-\frac{\uh^2}{R^2}\Big)\frac{u^4}{\uh^4},
\label{eq:metric}
\end{equation}
where $dH_3^2$ is the metric of the unit hyperbolic space. At the temperature $T=T_0/n$, with $T_0=1/2\pi R$ and $n$ the R\'enyi index, the horizon is at \mbox{$\uh(n)=4nR/(1+\sqrt{1+8n^2})$}. At $n=1$, the horizon is at $\uh=R$, so the metric is pure AdS$_5$ in AdS--Rindler coordinates. With $\Psi=\rho\,e^{i\theta}/\sqrt2$ and $B_a=A_a-\partial_a\theta$, the condensed phase of \cite{MainPaper} is
\begin{equation}
eL\rhob=\frac{p\,u^2}{1+\mu^2u^2/4},\qquad \Bb_t=\frac{\mu f}{1+\mu^2u^2/4},\qquad p^2=\frac12\big(\mu^2-\muc^2\big)\Big(\mu^2+\frac{4}{\uh^2}\Big),
\label{eq:family}
\end{equation}
with the critical chemical potential
\begin{equation}
\muc(n)^2R^2=\frac{1+\sqrt{1+8n^2}-4n^2}{2n^2},
\label{eq:mucn}
\end{equation}
which vanishes at $n=1$ and is negative for $n>1$. At $\mu=0$ the profile is exactly $eL\rhob=p\,u^2$, with $p^2=-2\muc^2/\uh^2$. In the probe sector the grand potential of condensation is $\Om_\SF-\Om_\N=-(L/8e^2)(\mu^2-\muc^2)^2$ per unit volume of $H^3$ \cite{MainPaper}. The combination $\Nn\equiv\pi^2L/e^2$ is proportional to the two-point coefficient of the current. The R\'enyi entropies follow from the free energy on $\mathbb{R}\times H^3$ at $T_0/n$ \cite{CHM2011,HMSY2011}.

We work in the conformal frame $S^1\times H^3$. Let the entangling surface $\Sigma$ be a plane, at $r=0$ in flat Euclidean space with polar coordinates $(r,\tau/R)$ in the transverse plane and $\vec y$ along $\Sigma$. The flat metric then reads
\begin{equation}
dr^2+\frac{r^2}{R^2}\,d\tau^2+d\vec y^{\,2}=\frac{r^2}{R^2}\Big[d\tau^2+R^2\,\frac{dr^2+d\vec y^{\,2}}{r^2}\Big].
\label{eq:frame}
\end{equation}
The bracket is $S^1\times H^3$, where $H^3$ has radius $R$ and Poincar\'e coordinate $z=r$. On the $n$-sheeted cover, $\tau\in[0,2\pi nR)$ and the circle has circumference $2\pi nR=1/T$; the entangling surface is the boundary of $H^3$, at $z\to0$. A primary of dimension $\Delta$ transforms as $O_{\rm flat}=(R/r)^\Delta O_{S^1\times H^3}$. For a spherical entangling surface the same holds with the ball model of $H^3$ \cite{CHM2011}.

\paragraph{Defect primaries and the two defects.}\label{sec:primaries} The conformal group of a planar surface defect, SO(3,1), acts as the isometry group of $H^3$, and transverse rotations act as translations along $S^1$. A defect primary is therefore a bulk mode that is homogeneous on the circle up to a phase and transforms as a field in AdS$_3=H^3$ \cite{GHJK2021,CuomoZhang2023}. Consider a Euclidean mode $\phi(u)\,e^{ik\tau/nR}\,Y(\vec y,z)$ on the black hole, with $-\nabla^2_{H^3}Y=\eta\,Y$ on the unit $H^3$, regular at the horizon and with no source at the AdS boundary. It is dual to a defect primary of transverse spin $k/n$, whose dimension $\Dh$ is fixed by
\begin{equation}
\Dh(\Dh-2)=-\eta,\qquad \Dh_\pm=1\pm\sqrt{1-\eta}.
\label{eq:dhat}
\end{equation}
Three cases matter. For $\eta=0$ the mode is constant on $H^3$ and $\Dh\in\{0,2\}$. A condensate that is constant on $H^3$ is the identity in the bulk-to-defect expansion of $O$, that is, a one-point function $\langle O\rangle=a_O/r^\Delta$, while a constant linear mode is the source of a defect operator of dimension two, which is marginal on a surface. For $\eta<1$ both roots are real; the one that reduces to $\Dh=\Delta$ at $n=1$, where the defect is trivial, is $\Dh_+$, and for $\eta<0$ it exceeds $2$, so the operator is irrelevant, while for $0<\eta<1$ it lies below $2$, so the operator is relevant. For $\eta\ge1$ the mode lies in the $L^2$ spectrum of the Laplacian on $H^3$, which starts at $\eta=1$, and $\Dh=1\pm i\nu$ with \mbox{$\nu=\sqrt{\eta-1}$}. A static, source-free mode in this range is the normalizable instability of \cite{BMM2013}: the static mode with $\eta=1$ of the normal phase at $\mu=0$ appears there at $n_*\approx6.58$, as also in \cite{MainPaper}.

We call $\Dz$ the R\'enyi defect of the normal phase, with normalizable boundary conditions on $H^3$, and $\Dsb$ the defect described by the homogeneous hairy saddle \eqref{eq:family}. Since the condensate of $\Dsb$ is constant on $H^3$, $\langle O\rangle\neq0$ near the entangling surface and decays as $r^{-2}$ away from it: $\Dsb$ is a conformal surface defect that breaks the U(1), not a state of $\Dz$.

\subsection{Marginality at $n=1$}
\label{sec:marginal}

In the rest of this section $\mu=0$, and in this subsection $R=1$.

At $n=1$ there is no defect, and the lowest operator of transverse spin zero in the bulk-to-defect expansion of $O$ is $O$ itself restricted to $\Sigma$, of dimension $\Dh=\Delta$. In $d$ dimensions eq.~\eqref{eq:dhat} becomes $\Dh(\Dh-d+2)=-\eta$, so $\eta=\Delta(d-2-\Delta)$. The static homogeneous mode is therefore a zero mode at $T_0$ precisely when $\Delta=d-2$, that is, when $\int_\Sigma O$ is marginal on the codimension-two entangling surface. In $d=4$ this is the BF bound, $\Delta=2$, which is why the homogeneous hairy phase sets in at exactly $n=1$ \cite{DMRS2010,BMM2013}, for any $\lambda>0$ \cite{MainPaper}. The corresponding mass, $m^2L^2=-2(d-2)$, at which the constant mode on the massless topological black hole has zero frequency, was identified in \cite{BelinMaloney2014}; we add its reading as a marginal defect operator.

The flow away from $n=1$ can be followed explicitly. At $\mu=0$ the field equation for $P=eL\rho$, with a dependence on $H^3$, is
\begin{equation}
u^5\partial_u\big(fu^{-3}\partial_uP\big)+u^2\nabla^2_{H^3}P+4P-\tfrac12P^3=0\qquad(R=1).
\label{eq:PH3}
\end{equation}
At $n=1$, $f=1-u^2$ and $P=u^2$ is an exact zero mode; for every $n$ the homogeneous solution is exactly $P=p\,u^2$. Projecting \eqref{eq:PH3} with $P=\sigma\,u^2$ onto $u^2$ gives
\begin{equation}
\nabla^2_{H^3}\sigma=M^2\sigma+g\,\sigma^3,\qquad M^2=-\frac{2(\uh^2-1)}{\uh^2},\qquad g=\frac{\uh^2}{4},
\label{eq:project}
\end{equation}
whose homogeneous solution, $\sigma^2=-M^2/g=8(\uh^2-1)/\uh^4$, is the exact $p^2$. In the variable $s=\ln(z/\delta)$, with cutoff $\delta$, $\nabla^2_{H^3}=\partial_s^2-2\partial_s$ on functions of $z$, and \eqref{eq:project} is a defect renormalization-group flow. At $n=1$, $M^2=0$ and $\sigma\simeq(g s)^{-1/2}$: the coupling of $\int_\Sigma O$ is marginally irrelevant for $\lambda>0$, and exactly marginal at $\lambda=0$ at infinite $\Nn$. For $n>1$, $M^2<0$, so the operator is relevant, and the flow ends at $\Dsb$. Since $u^2$ is the exact zero mode at $n=1$, the projection is exact to first order in $n-1$ (a Rayleigh quotient about an exact eigenfunction), which gives
\begin{equation}
\Dh_+=2-\tfrac23(n-1)+O\big((n-1)^2\big),\qquad \Dh_{\rm amp}=2+\tfrac43(n-1)+O\big((n-1)^2\big)
\label{eq:slopes}
\end{equation}
for the charged operator of $\Dz$ and the lowest amplitude operator of $\Dsb$. The slope of $\Dh_+$ is the general value $-\Delta/(d-1)$ for the defect operator of a scalar primary at transverse spin zero \cite{BFKW2017,LLMS2017}. The ratio of the two slopes, $-2$, is that of a weakly relevant fixed point with a cubic beta function. Near $n=1$, $\Dsb$ is thus a defect analog of a Wilson--Fisher fixed point, with $2-\Dh_+$ playing the role of the small parameter.

\subsection{Spectrum and conformality loss at $n_*$}
\label{sec:spectrum}

\begin{figure}[t]
\centering
\includegraphics[width=0.6\textwidth]{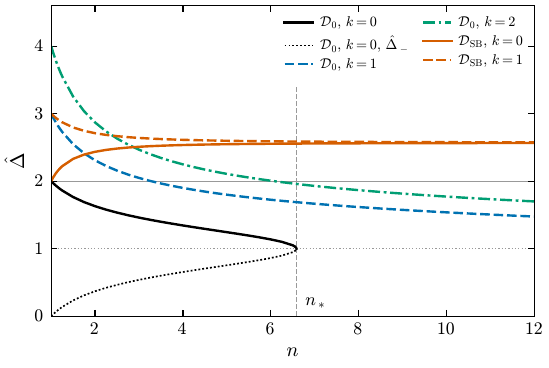}
\caption{The lowest defect primaries at $\mu=0$, of transverse spin $k/n$. The charged operator of $\Dz$ with $k=0$ is relevant for $1<n<n_*$ and meets its alternate-quantization partner at $\Dh=1$ at $n_*$; beyond $n_*$ its dimension becomes complex. The amplitude operators of $\Dsb$ are irrelevant. This and all later figures were plotted from the computed data with programs written by Claude Opus 5.5 (Anthropic).}
\label{fig:spectrum}
\end{figure}

The dimensions of the lowest defect primaries follow from \eqref{eq:dhat} once $\eta$ is known for each transverse spin. Writing the radial mode as $\phi=u^2(1-u^2/\uh^2)^{k/2}\chi(u)$ with $\chi$ smooth removes the source at the boundary and selects the regular branch at the horizon, which leaves a generalized eigenvalue problem in $\eta$. The results are in figure~\ref{fig:spectrum} and table~\ref{tab:spectrum}, and appendix~\ref{app:numerics} gives the method and its controls. All computed operators of $\Dsb$ other than the tilt are irrelevant. For the amplitude sector this follows from the positivity identity of \cite{MainPaper}: for a static mode the radial quadratic form is a sum of $\rhob^2(\partial_u(\delta\rho/\rhob))^2$ and $2\lambda\rhob^2\delta\rho^2$ with positive coefficients, which forces $\eta<0$ and hence $\Dh>2$ for the dual operator. In the limit $n\to\infty$, where the black hole is extremal, the radial equation with $k=0$ is hypergeometric. The mode of the amplitude operator with the largest $\eta$ is then $\phi=u^2(1-u^2/2R^2)^{\sqrt3-1}$, with $\eta=2-2\sqrt3$, so $\Dh=1+\sqrt{2\sqrt3-1}=2.570$, the limit of the computed amplitude dimensions.

For $1<n<n_*$ the R\'enyi defect $\Dz$ has a relevant charged operator with $k=0$. Those with $k=1$ and $2$, of transverse spin $1/n$ and $2/n$, become relevant at $n=3.28$ and $6.08$ and reach $\Dh=1$ at $n=31.1$ and $59.2$, beyond which their dimensions are complex. As $n\to\infty$ the Matsubara term, which only lowers $\eta$, vanishes away from the horizon, so in every channel of fixed $k$ the largest $\eta$ tends to $\frac32$, at which the AdS$_2$ throat at $n=\infty$ reaches its BF bound and $\Dh=1\pm i/\sqrt2$. Each such channel therefore reaches $\Dh=1$ at a finite index. The defect $\Dz$ has no relevant neutral operator from this sector, since the square $|\hat O|^2$ of the charged operator has dimension $2\Dh_+>2$ at large $\Nn$. The R\'enyi defect is therefore stable against small deformations that preserve the U(1), and reaching $\Dsb$ from it requires one that breaks the symmetry, a pinning field $h\int_\Sigma\hat O$ on the entangling surface. In the frame \eqref{eq:frame} such a deformation is a source for the mode $z^{2-\Dh_+}$ at $\partial H^3$, the nonnormalizable falloff of \cite{BMM2013}. The homogeneous saddle computes R\'enyi entropies when operators at the entangling surface source it \cite{BMM2013}, and at $\mu=0$ for $n<n_*$ only then \cite{MainPaper}. Below $n_*$, $\Dsb$, which we then call the pinned defect, is the infrared fixed point of $\Dz$ deformed by a pinning field, and section~\ref{sec:pinning} shows the flow.

At $n_*$ the two roots $\Dh_\pm$ meet at $\Dh=1$, the bottom of the spectrum on $H^3$, and beyond $n_*$ they are complex. The normalizable onset of \cite{BMM2013} is therefore a merger of $\Dz$ with its alternate-quantization partner, the defect analog of a fixed-point merger and conformality loss \cite{KLSS2009,GRZ2018}, as for Wilson lines above a critical charge \cite{ACKMR2022,ACKMR2023}. The same happens on the conical singularity of the O($N$) model in the $\varepsilon$-expansion: the two fixed points of the coupling of $\phi^2$ there, with $\phi$ the O($N$) field, merge and disappear beyond a critical index, and the flow runs away, which was read as a tendency toward $\langle\phi\rangle\neq0$ \cite{MFS2009}. Since such a condensate would be less than two-dimensional, the large-$N$ analysis there suggested instead a scale-invariant endpoint without condensation.

At large $\Nn$ the two fixed points of a double-trace coupling $|\hat O|^2$ on the defect are the two quantizations $\Dh_\pm$, between which the coupling flows \cite{Witten2001,Papadimitriou2007}. When a bulk scalar violates the BF bound, the fixed points of the double-trace coupling become complex, with dimensions of real part $\frac d2$ \cite{PR2008}; on the defect the bound is $\eta=1$ and the dimensions are $1\pm i\nu$. Beyond $n_*$ the double-trace coupling has no real fixed point. A conformal defect of a unitary theory, as at integer $n$, cannot have complex dimensions, and at any real $n$ the charged mode lowers the action, so $\Dz$ is not a consistent defect at large $\Nn$. This is conformality loss.

\section{The U(1)-breaking defect}
\label{sec:defect}

The defect $\Dsb$ is where the flows of section~\ref{sec:flow} end, and this section gives its data in closed form. Here $\mu=0$, except in eq.~\eqref{eq:db} and section~\ref{sec:tilt}.

\subsection{One-point coefficient}
\label{sec:onepoint}

The condensate is $\langle O\rangle=(L^{1/2}/e)\,p/\sqrt2$ on $H^3$ \cite{MainPaper}. By the Weyl factor of \eqref{eq:frame}, the one-point function at the distance $r$ from the entangling surface is $\langle O(x)\rangle=a_O/r^2$ with $a_O=R^2\langle O\rangle$.

In the normalization of \cite{MainPaper}, with $u_c\to0$ an ultraviolet cutoff in $u$ and $\Psi^{(b)}$ the source of $O$, the scalar behaves near the boundary as $\Psi=(u^2/L^{3/2})[-\Psi^{(b)*}\ln(u/u_c)+\langle O\rangle]$. Holographic renormalization is treated in \cite{deHaro2001,BFS2002,Skenderis2002}, and the logarithmic modes at the BF bound in \cite{Papadimitriou2007}. The regular solution in Euclidean AdS$_5$ at momentum $k$ is $\Psi\propto u^2K_0(ku)$, with $K_0(ku)=-\ln(ku/2)-\gamma_E+O(u^2\ln u)$, so $\langle O\rangle_k=\Psi^{(b)*}_k\,(-\ln k+{\rm const})$. The four-dimensional Fourier transform of $-\ln k$ is $1/(2\pi^2x^4)$ at $x\neq0$. Hence $\langle O(x)\,O^\dagger(0)\rangle=C_O/x^4$ with $C_O=1/2\pi^2$. It is independent of $L$ and $e$, since the scalar is canonically normalized in the bulk. For a unit-normalized operator,
\begin{equation}
\frac{a_O}{\sqrt{C_O}}=\sqrt{\Nn}\,R^2p,\qquad \frac{a_O^2}{C_O}\Big|_{\mu=0}=\Nn\,\frac{\big(4n^2-1-\sqrt{1+8n^2}\big)\big(1+\sqrt{1+8n^2}\big)^2}{16n^4}.
\label{eq:aO}
\end{equation}
The ratio $a_O^2/C_O$ vanishes at $n=1$, grows as $\frac{16}{3}\Nn\,(n-1)+O((n-1)^2)$ and tends to $2\Nn$ as $n\to\infty$. So $a_O/\sqrt{C_O}$ is of order $\sqrt\Nn$, as for a classical saddle of an action of order $\Nn$.

\subsection{Anomaly coefficients, conformal weight and the $b$-theorem}
\label{sec:b}

A conformal surface defect has a Weyl anomaly whose Euler term has the coefficient $b$, and, with $b$ normalized as in \cite{JensenOBannon2015}, a spherical defect of radius $R$ has $\ln\langle\mathcal{D}\rangle\supset(b/3)\ln(R/\delta)$. Along defect renormalization-group flows $b$ does not increase \cite{JensenOBannon2015}. The regulated volume of $H^3$ has the universal part $V_{H^3}\supset-2\pi R^3\ln(2R/\delta)$ \cite{MainPaper}, so at fixed $n$ the two defects differ by
\begin{equation}
\Delta b\equiv b(\Dsb)-b(\Dz)=6\pi R^3\beta\big(\Om_\SF-\Om_\N\big)=-\frac32\,\Nn\,nR^4\big(\muc^2-\mu^2\big)^2,
\label{eq:db}
\end{equation}
with $\beta=2\pi nR$. At $\mu=0$, $\Delta b=-\frac32\Nn\,n\,\muc^4R^4=-\frac{32}{3}\Nn(n-1)^2+O\big((n-1)^3\big)$. It is negative: a flow from $\Dz$ to $\Dsb$ is allowed by the $b$-theorem and the reverse is not. The sign comes from the negative universal part of $V_{H^3}$, through which the lower free energy of the hairy saddle becomes a lower $b$. That $\Delta b$ is quadratic in $n-1$ fits a weakly relevant flow with a quartic coupling, for which $\Delta b\propto(2-\Dh_+)^2/g$, with $g$ the coupling of eq.~\eqref{eq:project}.

The one-point function of the stress tensor defines the conformal weight $h_n$: near a planar conformal surface defect at index $n$, $\langle T_{ab}\rangle=-(h_n/2\pi)\,\delta_{ab}/r^4$ in the directions along it, at the distance $r$, with the stress tensor summed over the $n$ replicas. For a R\'enyi defect $h_n=\frac{2\pi n}{3}R^4\big[\ee(T_0)-\ee(T_0/n)\big]$, with $\ee$ the energy density on $H^3$ \cite{HMS2014}. The derivation rests only on the conformal map to $S^1\times H^3$, so it applies to $\Dsb$, whose stress tensor there is homogeneous and traceless. At order $\Nn$ the energy density follows from the grand potential of the probe sector, $\ee=\partial_\beta(\beta\Om)$ at $\mu=0$, as the entropy does in \cite{MainPaper}, where the action of a source-free scalar at the BF bound is shown to have no divergences. A change of $\beta$ is a change of the boundary metric, so this is the energy density of the boundary stress tensor at linear order in the backreaction (appendix~\ref{app:backreaction}). With $\beta\partial_\beta=n\partial_n$, and $\ee(T_0)$ common to both defects,
\begin{equation}
\Delta h_n\equiv h_n(\Dsb)-h_n(\Dz)=-\frac{2\pi n}{3}R^4\,\partial_n\big[n\big(\Om_\SF-\Om_\N\big)\big]=\frac{n\Nn}{12\pi}\,\frac{d}{dn}\big(n\muc^4R^4\big)=-\frac{n}{18\pi}\,\partial_n\Delta b
\label{eq:dh}
\end{equation}
at $\mu=0$. It is $\frac{32}{27\pi}\Nn(n-1)+O\big((n-1)^2\big)$ near $n=1$, $0.70\,\Nn$ at $n_*$ and $n\Nn/3\pi$ at large $n$, and it is positive for every $n>1$, since $\muc^2$ is negative and decreases with $n$.

Besides the Euler term, the Weyl anomaly of a surface defect in four dimensions has two terms, built from the traceless second fundamental form and from the pullback of the Weyl tensor, with coefficients $d_1$ and $d_2$ \cite{JORR2018}. In that normalization, $b$ is the one used above. The conformal weight fixes $d_2=18\pi h_n$ \cite{JORR2018} and, for a R\'enyi defect, the coefficient of the Weyl tensor in the logarithmic term of the R\'enyi entropy, $f_c(n)=\frac{3\pi h_n}{2(n-1)}$ \cite{LP2014}. The last form of \eqref{eq:dh}, $\Delta d_2=-n\,\partial_n\Delta b$, is the relation of \cite{LP2014} between $f_c$ and the coefficient $f_a$ of the intrinsic curvature, applied to the two defects, with $b=12(n-1)f_a$ and $d_2=12(n-1)f_c$. From $\Dz$ to $\Dsb$, then, $b$ decreases while $h_n$ and $d_2$ increase. The conformal weight in a hairy hyperbolic phase was obtained before in \cite{Sun2023}, from the mass of the neutral hairy black holes of \cite{Ren2019} in AdS$_4$, which include backreaction.

\subsection{Tilt operator}
\label{sec:tilt}

Since $\Dsb$ breaks the U(1), the divergence of the current has a contact term on the defect, $\partial\cdot J=\delta^2(x_\perp)\,\tilt(\vec y)$. This defines the tilt operator $\tilt$, of dimension exactly two \cite{CuomoZhang2023}. In the bulk it is the Goldstone mode of the condensate: a static gradient mode $b_i=-g(u)\,\partial_i\theta_0$, with $\theta_0$ harmonic on $H^3$. Its profile $g$ obeys the London equation
\begin{equation}
\partial_u\Big(\frac{f}{u}\,\partial_ug\Big)=\frac{(eL\rhob)^2}{u^3}\,g,\qquad g(0)=1.
\end{equation}
On $H^3$ this has the rational solution
\begin{equation}
g=1-\frac{c\,u^2}{1+\mu^2u^2/4},\qquad c=\frac{p^2}{2(\mu^2+2/R^2)}.
\label{eq:London}
\end{equation}
The same form, with the curvature of the slicing in place of $-1/R^2$, solves the London equation on every slicing of \cite{MainPaper}. On the plane it gives the superfluid density found there.

The Euclidean action of a slowly varying phase on $H^3$ is $(K/2)\int_{H^3}(\partial\theta_0)^2$, with $K=\beta\rho_s$ and the stiffness $\rho_s=2cL/e^2$. Take the unit $H^3$, $ds^2=(dz^2+d\vec y^{\,2})/z^2$. There the phase is a massless field with Dirichlet data $\theta_0\to\theta_\partial(\vec y)$ at $z\to0$. The boundary value $\theta_\partial$ is the local rotation angle of the defect, so it is the source of $\tilt$, with unit coefficient. At momentum $k$ along the defect the solution is $\theta_{0,k}=\theta_{\partial,k}\,kzK_1(kz)$, with $kzK_1(kz)=1+\frac12k^2z^2[\ln(kz/2)+\gamma_E-\frac12]+O(k^4z^4\ln kz)$. The on-shell action is therefore $-\frac K2\int\frac{d^2k}{(2\pi)^2}|\theta_{\partial,k}|^2k^2\ln k$ up to local terms, and $\langle\tilt\tilt\rangle_k=K\,k^2\ln k$. Since $\int\frac{d^2k}{(2\pi)^2}e^{ik\cdot y}k^2\ln k=2/(\pi|\vec y|^4)$ at $\vec y\neq0$, the two-point function is $\langle\tilt(\vec y)\tilt(0)\rangle=C_{\tilt}/|\vec y|^4$ with $C_{\tilt}=2K/\pi$. For radius $R$, $C_{\tilt}=2KR/\pi$, and so
\begin{equation}
C_{\tilt}=4nR^2\rho_s,\qquad C_{\tilt}\big|_{\mu=0}=\frac{2n}{\pi^2}\,\frac{a_O^2}{C_O}=\frac{32}{3\pi^2}\,\Nn\,(n-1)+O\big((n-1)^2\big).
\label{eq:Ct}
\end{equation}

Let $b_{O\tilt}$ be the coefficient of $\tilt$ in the bulk-to-defect expansion of $O$. In the frame \eqref{eq:frame}, with $R=1$, the bulk-to-defect two-point function of $O$ and $\tilt$ is $\langle O_{H^3}(z,\vec y)\,\tilt(0)\rangle=b_{O\tilt}C_{\tilt}\,z^2/(z^2+\vec y^{\,2})^2$. A local rotation of the defect by $\theta_\partial$ rotates the condensate, $\delta\langle O\rangle=i\langle O\rangle\theta_0$. Here $\theta_0$ is $\theta_\partial$ propagated by $\pi^{-1}z^2/(z^2+\vec y^{\,2})^2$, the bulk-to-boundary propagator of a field with $\Dh=2$ on $H^3$. Comparing the two gives $i\,a_O=\pi\,b_{O\tilt}\,C_{\tilt}$. This is the Ward identity of \cite{CuomoZhang2023} for a surface defect, which confirms that the normalization of $\tilt$ is theirs.

The proportionality of $C_{\tilt}$ to $a_O^2$ is special to $\lambda=e^2/2$, like the proportionality of the superfluid density to $\langle O\rangle^2$ on the plane \cite{MainPaper}. For other couplings, section~\ref{sec:lambda} gives how $C_{\tilt}$ depends on $\lambda$, with values in appendix~\ref{app:numerics}.

\section{Flows to symmetry breaking}
\label{sec:flow}

This section solves the flows from the R\'enyi defect to $\Dsb$, with a pinning field below $n_*$ and without one beyond it, and derives the universal R\'enyi term. Throughout we take $\mu=0$, the probe limit and the leading order in $1/\Nn$, and we set $R=1$ except in ratios such as $R/\delta$.

\subsection{Flows with a cutoff}
\label{sec:cutoff}

For a planar entangling surface, consider static configurations that depend on the radial coordinate $u$ and on the Poincar\'e coordinate $z$ of $H^3$, with $z\to0$ the entangling surface. With $P=eL\rho=u^2\chi(u,s)$ and $s=\ln(z/\delta)$, eq.~\eqref{eq:PH3} becomes
\begin{equation}
\mathcal{L}_u\chi+u\,\big(\partial_s^2-2\partial_s\big)\chi-\tfrac12u^3\chi^3=0,
\label{eq:flowpde}
\end{equation}
where $\mathcal{L}_u$ is the radial operator of the normal phase divided by $u^3$, so that $\mathcal{L}_u p=\frac12u^3p^3$ for the hairy saddle $\chi=p$. The cutoff at a distance $\delta$ from the entangling surface sits at $s=0$, and we impose the conditions
\begin{equation}
\chi(u,0)=\chi_{\rm UV},\qquad \chi(u,S)=p,
\label{eq:flowbc}
\end{equation}
with $S$ large. A positive $\chi_{\rm UV}$ is a pinning field at the cutoff, in units of the cutoff; $\chi_{\rm UV}=0$ means no source. The condition at $s=S$ fixes the infrared to $\Dsb$ only so that solutions can be found by Newton's method; the results below do not depend on $S$, and on the sphere (section~\ref{sec:sphere}) no infrared condition is imposed at all. A solution with a small source or none has a layer next to the cutoff, where $\chi$ is small, as in $\Dz$, joined at a wall to an interior where $\chi$ is close to $p$, as in $\Dsb$ (figure~\ref{fig:flow}(a)). We characterize a solution by the position $s_w$ of its wall, where $\chi(\uh/2,s)=p/2$. The $\langle O\rangle$ wall, where the condensate $\langle O\rangle$, proportional to $\chi(0,s)$, is half its interior value, lies farther out, by $0.25$ to $0.29$ for $n_*\le n\le30$.

Two numerical methods, Chebyshev collocation in $u$ and $s$ and fourth-order finite differences, solve \eqref{eq:flowpde}--\eqref{eq:flowbc}; appendix~\ref{app:numerics} describes them and gives their controls.

\paragraph{Why the endpoint is homogeneous.} Since the conformal group of the defect acts by isometries of $H^3$ and translations along $S^1$ (section~\ref{sec:primaries}), a conformal defect is described by a saddle that is invariant under these transformations up to gauge transformations. The gauge-invariant fields $|\Psi|$ and $B_a$ are then invariant under the identity component of SO(3,1), which acts on $H^3$ transitively with isotropy SO(3). So they are homogeneous on $H^3$, and $B_a$ has no component along it. Only the transverse rotations can be combined with the U(1), as for the defects of section~\ref{sec:flux}. If the flow from $\Dz$ ends at a conformal defect, its endpoint is therefore a homogeneous saddle, and any dependence on the position in $H^3$ is the flow itself, through the distance $z$ from the entangling surface. Among the homogeneous saddles at $\mu=0$, $\Dz$ is not a consistent defect beyond $n_*$, and $\Dsb$ is nodeless, which makes it linearly stable and the minimum among states of finite energy relative to it \cite{MainPaper}. On the cutoff ball beyond $n_*$ the source-free flow does end at $\Dsb$ as the ball grows (the proposition of section~\ref{sec:sphere}). At finite $\Nn$ the endpoint is not strictly a conformal defect (section~\ref{sec:finiteN}).

\paragraph{Pinning below $n_*$.}\label{sec:pinning} At $n=4$, as the pinning field is lowered, the wall moves into the infrared at the rate $ds_w/d\ln(p/\chi_{\rm UV})=1/(2-\Dh_+)$ that a pinning flow driven by the charged operator of $\Dz$ predicts. As the pinning is removed the wall runs off to the infrared and $\Dz$ fills the defect: below $n_*$ the R\'enyi defect is $\Dz$, and $\Dsb$ is the infrared end of the pinning flow, as anticipated in section~\ref{sec:spectrum}. Figure~\ref{fig:flow}(c) shows the wall position.

\subsection{Source-free flows beyond $n_*$}
\label{sec:sourcefree}

\begin{figure}[t]
\centering
\includegraphics[width=\textwidth]{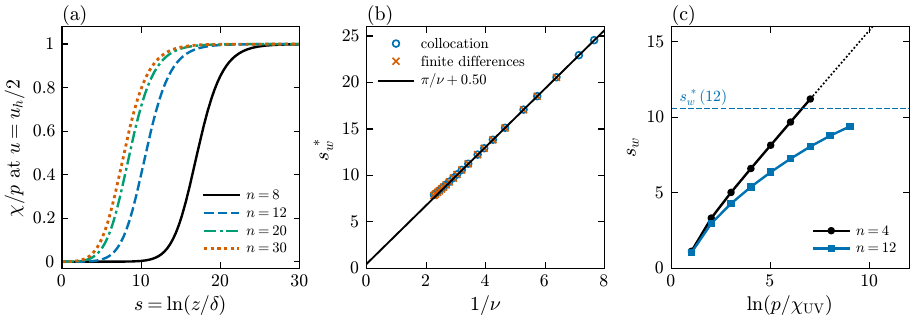}
\caption{Flows from the R\'enyi defect. (a) Source-free flows ($\chi=0$ at the cutoff) to the hairy saddle beyond $n_*$, at $u=\uh/2$ (finite differences). (b) The wall position of the lowest branch against $1/\nu$ (both methods) and the line $s_w^*=\pi/\nu+0.50$; the asymptotic law \eqref{eq:bkt} has $c_0=0.518$ for this definition of the wall. (c) The wall with a pinning field $\chi_{\rm UV}$ at the cutoff: at $n=4$ it runs off with the slope $1/(2-\Dh_+)$; at $n=12$ it approaches the source-free value from below.}
\label{fig:flow}
\end{figure}

For $7.5\le n\le30$ both methods find the same solution with $\chi_{\rm UV}=0$, whose wall sits at a finite distance $s_w^*$ from the cutoff (figure~\ref{fig:flow}(a) and table~\ref{tab:walls}) and does not depend on $S$. A separate computation, with two further solvers on the ball model of $H^3$ and only regularity imposed at the center, confirms these results (appendix~\ref{app:numerics}). It finds the nodeless source-free solution for $6.59\le n\le30$, closer to $n_*$ than the collocation and finite-difference methods reach.

With $\chi=0$ at the cutoff and only boundedness in the interior, no nonzero solution can exist for $n\le n_*$. The field equation, multiplied by the field and integrated, equates a quadratic form, nonnegative for $n\le n_*$ because the Dirichlet spectrum of $-\nabla^2$ on the cutoff region lies at or above $1$, to minus a positive quartic integral, so $\chi=0$, as in \cite{MainPaper}. Conversely, a positive solution exists for every $n>n_*$ once the cutoff ball is large enough (the proposition of section~\ref{sec:sphere}). The remark of \cite{BMM2013} that the constant mode is not normalizable on $H^3$ and requires a source stays true for the homogeneous saddle taken by itself; beyond $n_*$, however, the saddle is reached as the infrared end of these flows rather than imposed by a source. The breaking of the U(1) beyond $n_*$ is therefore spontaneous, at large $\Nn$, and the phase expected there to break the symmetries of hyperbolic space \cite{BMM2013,BHMM2014} is the homogeneous saddle beyond a layer at the entangling surface.

The solutions come in branches. The one above is the lowest and has no interior zero. Both methods find a second branch whose profile has one interior zero and whose wall lies farther out by $\pi/\nu$. On the plane the free energy of any source-free flow per unit area of the entangling surface, measured from the normal phase, is exactly $-(L/4e^2\delta^2)\int du\,u\,(\partial_s\chi)^2|_{s=0}$ (section~\ref{sec:endpoint}), which on the sphere gives the area term of section~\ref{sec:sphere}. It is therefore negative, and most negative on the lowest branch: on the second, whose slope at the cutoff is $e^{-\pi/\nu}$ times that of the lowest, it is $e^{-2\pi/\nu}$ times as large (appendix~\ref{app:numerics}).

\paragraph{Stability.}\label{sec:stability} Static perturbations of a flow solution with Dirichlet data at both ends of the $s$-interval are governed by $J=\mathcal{L}_u+u(\partial_s^2-2\partial_s)-\frac32u^3\chi^2$, and a positive eigenvalue of $J$, with respect to a positive weight, signals an instability. The scalar part of the operator of phase fluctuations is $J_\theta=\mathcal{L}_u+u(\partial_s^2-2\partial_s)-\frac12u^3\chi^2$, and $J_\theta\chi=0$ by \eqref{eq:flowpde}. Since $\chi$ is positive in the interior and nonnegative at both ends, $J_\theta$ has no positive eigenvalue, and $J=J_\theta-u^3\chi^2$ has only negative ones; on the cutoff ball this is part of the proposition of section~\ref{sec:sphere}. The lowest branch is therefore stable against static perturbations in the amplitude sector, and at $\mu=0$ the phase sector has positive energy by the argument of \cite{MainPaper}. The softest mode lives in the layer, whose length is just below $\pi/\nu$, and its eigenvalue goes to zero as $n\to n_*$.

At $\mu=0$ static stability also excludes linear perturbations that grow in time on $\mathbb{R}\times H^3$. The background gauge field vanishes, so the energy of a perturbation is a positive kinetic term plus the static quadratic forms: that of $-J$ for the amplitude, and a positive one for the phase sector. As in \cite{MainPaper}, a growing mode would then carry a finite energy, conserved with Dirichlet data at the cutoff, and must vanish. The least-damped quasinormal modes of the flows are purely damped, at $\omega R=-0.0012i$, $-0.0045i$ and $-0.0069i$ for $n=8$, $12$ and $20$, respectively, on the planar domain with $S=40$.

\subsection{The \texorpdfstring{$\pi/\nu$}{pi/nu} law}
\label{sec:bkt}

Between the cutoff and the wall the field is small and obeys the linearization of \eqref{eq:flowpde} about $\Dz$. Its nodeless radial mode has $\eta=1+\nu^2$, so that $\chi\propto e^s\sin\nu(s-s_0)$, and the condition at the cutoff gives $s_0=0$. On the other side, the wall is a solution of the nonlinear problem whose ultraviolet tail, at $\nu\to0$, is proportional to $e^{t}(t+\ell)$ with $t=s-s_w$. Matching the two in the overlap region gives $\nu\cot(\nu s_w)=1/\ell$. For small $\nu$ the profile can stay positive only if $\nu s_w\to\pi$, and then the wall lies at
\begin{equation}
s_w^*=\frac{\pi}{\nu}+c_0+O(\nu^2),
\label{eq:bkt}
\end{equation}
where $c_0=\ell$. A profile with $j$ interior zeros has $\nu s_w\to(j+1)\pi$ instead, so the solutions form a tower whose walls are spaced by $\pi/\nu$, the discrete scale invariance that comes with complex dimensions; the second branch of section~\ref{sec:sourcefree} is its first excited member. A tower of condensed states that arises when two fixed points annihilate was found in a holographic system at finite density in \cite{ILM2011}. The numerical flows bear out eq.~\eqref{eq:bkt}: between the cutoff and the wall the profile is $e^s\sin(\nu s)$, and the coefficient of $1/\nu$ is $\pi$ (figure~\ref{fig:flow}(b) and appendix~\ref{app:numerics}).

The constant $c_0$ is not universal: it is $0.518$ for our wall and $0.769$ for the $\langle O\rangle$ wall, and it depends on the cutoff prescription. With the Robin condition $\partial_s\chi=\Gamma\chi$ at the cutoff in place of $\chi=0$, a double-trace coupling of $\Dz$ at the cutoff scale, the zero of the mode moves for $\Gamma>1$ to $s_0=-\arctan\big(\nu/(\Gamma-1)\big)/\nu$, and the wall moves with it. The computed flows follow this: the coefficient of $1/\nu$ does not change, and $c_0$ moves by $-1/(\Gamma-1)$ in the limit $\nu\to0$. For $\Gamma<1$ the operator $-\nabla^2$ on $H^3$ with this condition at the cutoff has a bound state at $1-(1-\Gamma)^2$, so the normal phase is unstable once $\eta_0$ exceeds this value, already below $n_*$: from $n=2.72$ at $\Gamma=\frac12$ and at every $n>1$ for the Neumann condition, which the homogeneous saddle satisfies. As for the attractive quartic of section~\ref{sec:pinu}, symmetry breaking then comes before conformality loss, so the transition at $n_*$ assumes a cutoff that stays in the basin of $\Dz$ up to $n_*$, such as $\chi=0$ or the Robin condition with $\Gamma>1$.

From the spectrum, $\nu^2/(n-n_*)\to0.0299$ as $n\to n_*$. Near $n_*$, therefore, $\nu\simeq0.173\sqrt{n-n_*}$, and the scale at which the R\'enyi defect crosses over from $\Dz$ to $\Dsb$ is
\begin{equation}
\frac{z_*}{\delta}=e^{s_w^*}\simeq e^{c_0}\exp\Big(\frac{18.2}{\sqrt{n-n_*}}\Big).
\label{eq:zstar}
\end{equation}
This is the scale of conformality loss \cite{KLSS2009}, of BKT type, here on a defect. In holography the same exponential scale appears when a scalar violates the BF bound of an AdS$_2$ region at finite density \cite{JKST2010,FHR2010,ILM2011}; here $H^3$ plays that role, with the bound at $\eta=1$. The separate computation follows the flows down to $n=6.59$, where $z_*/\delta\approx e^{244}$; away from $n_*$, $\pi/\nu(n)$ itself should be used in place of eq.~\eqref{eq:zstar}. The flow is correspondingly fragile against explicit breaking. A pinning field $\chi_{\rm UV}$ at the cutoff moves the wall by $0.42$ to $0.51$ of a unit of $s$ for $8\le n\le30$ once $\chi_{\rm UV}/p$ reaches $\delta/z_*=e^{-s_w^*}$, $4\times10^{-8}$ at $n=8$. Near $n_*$, explicit breaking of the U(1) at the entangling surface sets the crossover unless it is exponentially small, so the spontaneous scale $z_*$ needs a symmetric regulator, as in the replica construction.

\subsection{The sphere and the universal R\'enyi term}
\label{sec:sphere}

For a spherical entangling surface of radius $R$ we solve \eqref{eq:flowpde} on the ball model of $H^3$, with $\partial_s^2-2\partial_s$ replaced by $\partial_\zeta^2+2\coth\zeta\,\partial_\zeta$, $\zeta\in[0,\zeta_{\max}]$ and $\zeta_{\max}=\ln(2R/\delta)$. We require regularity at $\zeta=0$, set $\chi=0$ at the cutoff and impose \emph{no} condition in the interior. At $\mu=0$, $\mathcal{L}_u=\partial_u(uf\partial_u)+\frac12p^2u^3$, and the equation is the stationarity condition of
\begin{equation}
E[\chi]=4\pi\int_0^{\zeta_{\max}}d\zeta\,\sinh^2\zeta\int_0^{\uh}du\,\Big[uf(\partial_u\chi)^2+u(\partial_\zeta\chi)^2+\tfrac14u^3\big(\chi^2-p^2\big)^2-\tfrac14u^3p^4\Big],
\label{eq:Eball}
\end{equation}
where $(L/2e^2)E$ is the free energy of the probe sector per unit time for real $\chi$ with no source, regular at the horizon and zero at the cutoff. Let $\eta_0(n)$ be the largest $\eta$ of the radial operator, that of its nodeless mode, so that $\eta_0=1+\nu^2$ beyond $n_*$, and $\lambda_1=1+\pi^2/\zeta_{\max}^2$ the lowest Dirichlet eigenvalue of $-\nabla^2$ on the ball, with eigenfunction $\sin(\pi\zeta/\zeta_{\max})/\sinh\zeta$.

\paragraph{Proposition.} At $\mu=0$, in the probe limit and at leading order in $1/\Nn$, consider the real, source-free solutions on the cutoff ball that vanish at the cutoff and are regular at its center and at the horizon. Then: (i) $\eta_0(n)$ increases strictly with $n$, and whenever $\eta_0(n)\le\lambda_1$, in particular for every $n\le n_*$, the only such solution is $\chi=0$. For $n>n_*$ with $\nu(n)>\pi/\zeta_{\max}$: (ii) there is a solution with $0<\chi<p$ everywhere inside the cutoff, the AdS boundary and the horizon included; (iii) it is the only positive solution, and its stability operator is negative definite; (iv) it minimizes $E$ and, up to a global phase and a gauge transformation, the action of the probe sector; (v) as $\zeta_{\max}\to\infty$ it tends in the interior to the saddle $\chi=p$ of $\Dsb$, with $p-\chi$ decaying exponentially in the distance from the cutoff at every rate below $\Dh_{\rm amp}-2$, uniformly in $\zeta_{\max}$. In this limit a source-free flow exists exactly for $n>n_*$, as the minimum of this action.

The proof is in appendix~\ref{app:proof}: a change of variable turns the degenerate radial operator into a uniformly elliptic operator on a two-sphere, with the AdS boundary and the horizon at its poles, so that the argument of Brezis and Oswald for sublinear equations \cite{BrezisOswald1986} carries over.

\begin{figure}[t]
\centering
\includegraphics[width=\textwidth]{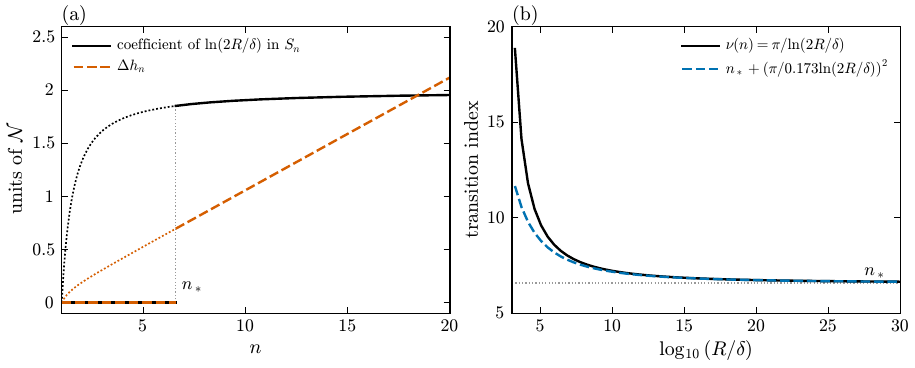}
\caption{The R\'enyi defect across $n_*$ at $\mu=0$, in the probe sector and at leading order in $1/\Nn$. (a) The coefficient of $\ln(2R/\delta)$ in $S_n$ and the change $\Delta h_n$ of the conformal weight, with $R/\delta\to\infty$ first: zero below $n_*$, where the R\'enyi defect is $\Dz$, and those of $\Dsb$, from eqs.~\eqref{eq:Snuniv} and \eqref{eq:dh}, beyond it. Dotted: $\Dsb$ below $n_*$, reached only with a pinning field. (b) The index at which the transition sits on a ball with finite $R/\delta$, where $\nu(n)=\pi/\ln(2R/\delta)$ (solid), and the approximation near $n_*$ (dashed).}
\label{fig:renyi}
\end{figure}

The proposition does not locate the wall of the flow; the numerical solutions by collocation and finite differences do (appendix~\ref{app:numerics}). The distance of their wall from the cutoff equals that of the planar flow, $s_w^*$.

On a solution the free energy $(L/2e^2)E$ is $4\pi\int d\zeta\,\sinh^2\zeta\;\Om(\zeta)$, with
\begin{equation*}
\Om(\zeta)=-\frac{L}{8e^2}\int du\,u^3\chi^4,
\end{equation*}
which reproduces $\Om_{\rm hom}\equiv\Om_\SF-\Om_\N$ for $\chi=p$ \cite{MainPaper}. Let $\Lambda(s)=\Om(\zeta_{\max}-s)-\Om_{\rm hom}$, with $s$ the distance from the cutoff, as on the plane. Since $4\sinh^2\zeta=e^{2\zeta}-2+e^{-2\zeta}$,
\begin{gather}
4\pi\int_0^{\zeta_{\max}}d\zeta\,\sinh^2\zeta\;\Om=\Om_{\rm hom}V_{H^3}(\zeta_{\max})+\pi e^{2\zeta_{\max}}I_2-2\pi I_0+O\big(e^{-0.55\,\zeta_{\max}}\big),\label{eq:layer}\\
I_k=\int_0^{\zeta_{\max}}ds\,e^{-ks}\Lambda(s).\nonumber
\end{gather}
The area term, the coefficient of $e^{2\zeta_{\max}}$, is $\pi(I_2-|\Om_{\rm hom}|/2)$. It vanishes for the normal phase in the probe sector at $\mu=0$, but not here: in units of the area term of the homogeneous saddle, $\pi\Om_{\rm hom}/2$, it is $2.9$--$4.3$ times $e^{-2s_w^*}$ for $7.5\le n\le30$, because the hairy region starts at the distance $z_*$ from the entangling surface. This term is exponentially small near $n_*$, where it behaves as $e^{-2\pi/\nu}$, and depends on the cutoff in any case.

For every $n>n_*$ the coefficient of $\ln(2R/\delta)$ in the free energy is the homogeneous one, $-2\pi\Om_{\rm hom}$: the planar flow moved onto the ball and the ball solution moved onto the plane are trial functions that bound the free energy from both sides (appendix~\ref{app:proof}). The layer changes only the constant term, which depends on the cutoff. With the Robin condition of section~\ref{sec:bkt} the coefficient of $\ln(2R/\delta)$ does not change; the area term is the same multiple of $e^{-2s_w^*}$, larger than with $\chi=0$ because the layer is shorter.

In defect language, for $R\gg z_*$ one has $\ln\langle\mathcal{D}\rangle\simeq\frac13b(\Dz)\ln(z_*/\delta)+\frac13b(\Dsb)\ln(R/z_*)$, and the coefficient of $\ln R$ in $\ln{\rm Tr}\,\rho_A^n$ is $b(\Dsb)/3$. Beyond $n_*$, at leading order in $1/\Nn$ and with $R/\delta\to\infty$ first, the universal term of the R\'enyi entropy is therefore the one of the homogeneous saddle at $\mu=0$ \cite{MainPaper}, without operators inserted at the entangling surface:
\begin{equation}
S_n\supset\frac{n}{2(n-1)}\,\Nn\,\muc^4R^4\,\ln\frac{2R}{\delta}=-\frac{\Delta b}{3(n-1)}\,\ln\frac{2R}{\delta}.
\label{eq:Snuniv}
\end{equation}
It jumps at $n_*$: the coefficient of the logarithm goes from zero (the probe sector of $\Dz$) to $1.85\,\Nn$, a change $\Delta b(n_*)=-31.1\,\Nn$ of the anomaly coefficient (figure~\ref{fig:renyi}(a)). As $\delta\to0$ the layer shrinks onto the entangling surface, so the conformal weight of the R\'enyi defect also jumps at $n_*$, by the value of eq.~\eqref{eq:dh}, $\Delta h_{n_*}=0.70\,\Nn$.

At finite $R/\delta$ the transition sits at the threshold $\nu(n)=\pi/\ln(2R/\delta)$ of the cutoff ball, $n=12.2$ for $R/\delta=10^4$ and $n=8.5$ for $R/\delta=10^6$, and it approaches $n_*$ only as $1/\ln^2(R/\delta)$ (figure~\ref{fig:renyi}(b)). In the O($N$) model near four dimensions the universal coefficients of the R\'enyi entropy also jump at a critical index, and the limits of large size and $n\to n_*$ do not commute, because a length scale diverges at the transition \cite{MFS2009}.

\section{How generic is the mechanism?}
\label{sec:generic}

Sections~\ref{sec:renyidefect}--\ref{sec:flow} use one model, a scalar at the BF bound with the Bogomolny quartic coupling, in which the saddles are in closed form. This section asks which parts of the mechanism depend on that choice. At $\mu=0$ and in the probe limit we vary the quartic coupling (section~\ref{sec:lambda}), the scalar mass (sections~\ref{sec:crossing} and~\ref{sec:endpoint}) and the shape of the potential (section~\ref{sec:pinu}). Table~\ref{tab:generic} summarizes the outcome.

\begin{table}[htb]
\centering
\small
\begin{tabular}{@{}>{\raggedright}p{0.33\textwidth} >{\raggedright}p{0.30\textwidth} >{\raggedright}p{0.29\textwidth}@{}}
\toprule
Part of the mechanism & Holds & Basis \tabularnewline
\midrule
Merger of $\Dh_\pm$ at $\Dh=1$ & given a crossing & kinematics, eq.~\eqref{eq:dhat} \tabularnewline
Crossing, $\eta_0(n)=1$ & for $1<\Delta<3$ & the AdS$_2$ throat \cite{BMM2013} \tabularnewline
Irrelevant amplitude operators & on any stable homogeneous saddle & the homogeneous action \tabularnewline
Homogeneous endpoint of the flow & proven in the model of sections~\ref{sec:renyidefect}--\ref{sec:flow}; argued for a probe action bounded below & symmetry, gradient flow, minimizer; flows at three $\Delta$ \tabularnewline
Transition at $n_*$, $\pi/\nu$ law & if no source-free flow at $n_*$ & matching, section~\ref{sec:bkt} \tabularnewline
Universal term of the saddle & given the endpoint; proven in the model & the layer integral; two-sided bound \tabularnewline
Closed forms; onset at $n=1$, $n_*\approx6.58$ & at the Bogomolny point; at the BF bound & model-specific \tabularnewline
\bottomrule
\end{tabular}
\caption{Which parts of the mechanism are generic, at $\mu=0$ and in the probe limit.}
\label{tab:generic}
\end{table}

\subsection{The quartic coupling}
\label{sec:lambda}

At $\mu=0$ the hairy saddle exists for every $\lambda>0$, with $(\lambda/e^2)p^2=-\muc^2/\uh^2$ \cite{MainPaper}. The spectrum of $\Dz$, the onset at $n=1$ and $n_*$ come from a linear problem and do not depend on $\lambda$. The amplitude operators of $\Dsb$ do not either, since their potential, $3(\lambda/e^2)P^2$, contains only $\lambda p^2$, and in terms of $\chi/p$ the same holds for the flows of section~\ref{sec:flow}. The one-point coefficient squared, $\Delta b$ and $\Delta h_n$ scale as $e^2/2\lambda$. The tilt normalization does not scale simply, since the London equation has the rational solution \eqref{eq:London} only at $\lambda=e^2/2$. Numerically it scales as $e^2/2\lambda$ for a weak condensate, near $n=1$ or at large $\lambda$, and as $(e^2/2\lambda)^{1/2}$ for a strong one, as the stiffness approaches $(L/e^2)p$ at small $\lambda$. The merger, the amplitude spectrum and the flows therefore do not need the Bogomolny coupling; the closed forms and the relation $C_{\tilt}\propto a_O^2$ do.

\subsection{The merger and the crossing window}
\label{sec:crossing}

The merger is kinematic. For any Euclidean bulk mode, whatever the field, the potential or the backreaction, the conformal group of the defect fixes $\Dh(\Dh-2)=-\eta$, eq.~\eqref{eq:dhat}, so the two quantizations meet at $\eta=1$, the bottom of the $L^2$ spectrum on $H^3$, and are complex beyond it. What depends on the model is whether the largest $\eta$ of the normal phase, $\eta_0(n)$ as in section~\ref{sec:sphere}, reaches $1$. Take a probe scalar of dimension $\Delta$ on the uncharged hyperbolic black hole, with $m^2L^2=\Delta(\Delta-4)$ and the alternate quantization for $\Delta<2$. With $P=u^\Delta\chi$, $x=u^2/\uh^2$ and the potential $\frac14c_3P^4+\frac16c_5P^6$ of the examples below, the field equation at $\mu=0$ and $R=1$ is
\begin{equation*}
\mathcal{L}_x\chi+\nabla^2_{H^3}\chi-c_3\uh^{2\Delta-2}x^{\Delta-1}\chi^3-c_5\uh^{4\Delta-2}x^{2\Delta-1}\chi^5=0,
\end{equation*}
where $\eta$ is an eigenvalue of the radial operator $\mathcal{L}_x$, which is self-adjoint with the weight $w=x^{\Delta-2}$; $c_3=\lambda/e^2$, and the model of section~\ref{sec:renyidefect} is $\Delta=2$, $c_3=\frac12$, $c_5=0$, where $\mathcal{L}_x=\mathcal{L}_u/u$. The weight does not depend on $n$ and the coefficients of $\mathcal{L}_x$ change monotonically with $n$, so $\eta_0(n)$ rises from $\Delta(2-\Delta)$ at $n=1$ and increases strictly, by the argument of appendix~\ref{app:proof}, part (i). For $\Delta<2$ both falloffs, $u^\Delta$ and $u^{4-\Delta}$, have finite energy, and the natural boundary condition of the Rayleigh quotient removes $u^{4-\Delta}$, the source of the alternate quantization. As $n\to\infty$ the black hole becomes extremal, with a throat AdS$_2\times H^3$ whose BF bound caps $\eta$, and $\eta_0$ approaches
\begin{equation}
\eta_\infty=\frac{-1-m^2L^2}{2}.
\label{eq:etainf}
\end{equation} Since $\eta_0(n)$ increases toward $\eta_\infty$, a merger exists if and only if $\eta_\infty>1$, that is, for $1<\Delta<3$. This is the window found in \cite{BMM2013} from the same throat, with the onset as a function of $\Delta$.

The root $n_*$ increases with $\Delta$, from $1.47$ at $\Delta=1.25$ to $52.7$ at $\Delta=2.5$ (appendix~\ref{app:numerics}), so an integer index lies beyond the merger once $\Delta$ is small enough: $n=2$ for $\Delta<1.438$, $n=3$ for $\Delta<1.658$ and $n=4$ for $\Delta<1.796$.

As $\Delta\to1$, $n_*\to1$, since at $n=1$ the operator has $\Dh=\Delta$ and sits at the defect BF bound; as $\Delta\to3$, $n_*$ diverges. For $3\le\Delta<2+\sqrt3$ the constant mode still condenses, since $0<\eta_\infty\le1$, but there is no merger at any $n$: the pinned defect exists and, with a repulsive potential, reaching it needs a source at every $n$, however large.

\subsection{The endpoint}
\label{sec:endpoint}

The argument of section~\ref{sec:cutoff} that a conformal endpoint of the flow is homogeneous uses only the conformal group. We argue from three further properties that the source-free flow reaches a homogeneous saddle for a scalar of any $\Delta$ with a potential $V(\rho)$, at $\mu=0$ and in the probe limit; the third also says which saddle. Write the flow equation as $\mathcal{L}_x\chi+(\partial_s^2-2\partial_s)\chi=F_\chi$, with $w$ the weight of section~\ref{sec:crossing} and $F$ the nonlinear part of the potential.
\begin{itemize}
\item \emph{The flow is gradient-like.} The function $H(s)=\int dx\,w\big[\frac12\chi_s^2+\frac12\chi \mathcal{L}_x\chi-F(\chi)\big]$ obeys $dH/ds=2\int dx\,w\,\chi_s^2\ge0$, so a bounded flow cannot cycle, and we expect it to approach an $s$-independent solution, a homogeneous saddle. This needs a Euclidean action that is convex in the gauge field. That holds at imaginary $\mu$, where the R\'enyi defect carries a flux (section~\ref{sec:flux}), and fails at real $\mu\neq0$, where $\Bb_\tau$ is imaginary and the action is concave in $B_t$. At real $\mu$ a different function decreases along the flows (section~\ref{sec:realmu}).
\item \emph{A minimizer exists.} If the action is bounded below, it has a minimizer on every cutoff ball, nonzero once $\nu(n)>\pi/\ln(2R/\delta)$, as in the proposition of section~\ref{sec:sphere}. A bulk potential bounded below is not enough: near the AdS boundary the gradient, quartic and sextic terms of the action scale alike, so a strong enough attractive quartic makes the action unbounded below there even when a sextic bounds $V(\rho)$. For the potential of section~\ref{sec:crossing} this happens for $c_3<-(\Delta-1)\sqrt{16c_5/3}$ (appendix~\ref{app:numerics}), below the value at which a domain wall that obeys a first-order equation has zero action. This is the probe limit of the superpotential condition for an energy bound in asymptotically AdS gravity, whose branch depends on the falloff \cite{AHHM2007}. The examples of section~\ref{sec:pinu} lie above this value.
\item \emph{It is unique.} If $V'(\rho)/\rho$ increases, the positive solution is unique by the argument of \cite{BrezisOswald1986}, given a bounded ratio of positive solutions at the AdS boundary, as at $\Delta=2$. In the interior it then approaches the unique positive homogeneous saddle, as appendix~\ref{app:proof} proves for the model of section~\ref{sec:renyidefect}. Otherwise the endpoint is still a stable homogeneous saddle, but which one, and where the transition lies, must be checked case by case (section~\ref{sec:pinu}).
\end{itemize}
At $\mu=0$ the function $H$ also gives the action on the plane, by a Pohozaev-type identity. For a bounded flow with $\chi=0$ at the cutoff, the action measured from the normal phase is, up to a constant factor, $\int ds\,e^{-2s}\int dx\,w\big[\frac12\chi_s^2-\frac12\chi \mathcal{L}_x\chi+F(\chi)\big]$, and on a solution its integrand in $s$ is the derivative of $\frac12e^{-2s}H$. So the action is $-\frac12H(0)=-\frac14\int dx\,w\,\chi_s^2\big|_{s=0}<0$, for any potential and on any branch.

Stability and irrelevance are equivalent here. The largest $\eta$ of the amplitude operator of transverse spin zero is positive, so that $\Dh<2$, if and only if a homogeneous static deformation lowers the action of the saddle. A homogeneous saddle that is a local minimum therefore has only irrelevant amplitude operators, and the positivity identity of section~\ref{sec:spectrum} is the case of an increasing $V'(\rho)/\rho$; at $\mu=0$ the Matsubara term of a nonzero transverse spin only adds positive energy. In a semiclassical limit of defects in scalar field theories, the defect operators that could destabilize a fixed point likewise stay irrelevant for potentials bounded below \cite{CBRR2023}. With several scalars the endpoint must also be stable in the channels of the others, as the analytic saddles of the top-down example of section~\ref{sec:discussion} are not.

Given the endpoint, the universal term is that of the homogeneous saddle: the on-shell action density approaches the saddle's at the rate of the leading irrelevant operator, so the layer integral $I_0$ of section~\ref{sec:sphere} converges.

The model with $\lambda=e^2/2$ at $\Delta=1.5$, $2.25$ and $2.5$, where the homogeneous saddle is not in closed form, confirms the argument (appendix~\ref{app:numerics}). The source-free flows are nodeless and positive, so the argument of section~\ref{sec:stability}, with $J_\theta=\mathcal{L}_x+\partial_s^2-2\partial_s-F_\chi/\chi$, makes them stable against static perturbations. The coefficient of $1/\nu$ in their walls is $\pi$, and the constant $c_0$ depends on $\Delta$. On the cutoff ball, with nothing imposed in the interior, the profile approaches the homogeneous saddle at the rate $\Dh_{\rm amp}-2$ and the coefficient of $\ln(2R/\delta)$ is the homogeneous one. At the onset of the hairy phase the slopes of $\Dh_{\rm amp}$ and of the charged operator of $\Dz$ have the ratio $-2$ at every $\Delta$ computed, as at $\Delta=2$ (section~\ref{sec:marginal}).

The flows at $\Delta=1.5$ include $n=3$, $4$ and $5$, and at $\Delta=1.25$ we find nodeless, positive source-free flows at $n=2$ and $3$. With $\lambda=e^2/2$ and $\Delta<1.438$, then, the universal terms of the second and third R\'enyi entropies are those of the homogeneous saddle, by the argument of this section. At $\Delta=1.25$ they are $S_2\supset18.0\,\Nn\ln(2R/\delta)$ and $S_3\supset14.2\,\Nn\ln(2R/\delta)$, against zero for the probe sector of $\Dz$, with $\Delta b=-54.1\,\Nn$ and $-85.0\,\Nn$, respectively.

We have not found an earlier argument, general or conditional, that a defect whose charged operator crosses the defect BF bound flows with no source to a homogeneous conformal defect that breaks the symmetry.

\subsection{When the \texorpdfstring{$\pi/\nu$}{pi/nu} law fails}
\label{sec:pinu}

An action bounded below is not enough for the transition at $n_*$ and its $\pi/\nu$ law. Let the solution at $n_*$ have an ultraviolet tail proportional to $e^t(t+\ell)$, as in section~\ref{sec:bkt} but with the origin of $t$ at a fixed distance $c$ from the wall, so that $\ell$ is the distance from that origin to the zero of the tail and matching gives $\nu\cot\big(\nu(s_w-c)\big)=1/\ell$. If $\ell$ is of order one or negative, the zero lies inside the wall or beyond it, matching gives eq.~\eqref{eq:bkt}, and no source-free flow exists at or below $n_*$, since the cutoff would have to sit at that zero. If $\ell\gg1$, the zero lies deep in the linear region, with the profile positive on its infrared side, and the cutoff can sit there. A source-free flow then exists already at $n_*$, with $s_w\simeq c+\ell$, and the walls beyond $n_*$ stay finite as $\nu\to0$. Below $n_*$ the linear region is $e^s\sinh\bar\nu s$ with $\bar\nu=\sqrt{1-\eta_0}$, matching gives $\bar\nu\coth\big(\bar\nu(s_w-c)\big)=1/\ell$, and flows exist down to $n_c=n_*-1/\big(\ell^2\eta_0'(n_*)\big)$, where the wall diverges logarithmically. The crossover scale then grows as a power, $z_*/\delta\propto(n-n_c)^{-\ell/2}$, not as $\exp(\pi/\nu)$. Within the matching argument, the $\pi/\nu$ law holds if and only if no source-free flow exists at $n_*$, which the energy argument of section~\ref{sec:sourcefree} guarantees whenever $\chi F_\chi\ge0$, as it is for quartic and sextic potentials with nonnegative coefficients.

An attractive quartic stabilized by a sextic can give $\ell\gg1$. At $\Delta=2$, with $c_3=-\frac12$ and $c_5=\frac1{10}$, the walls beyond $n_*$ lie far below $\pi/\nu$, at about half of it for $\nu=0.08$. Fitted to the walls just beyond $n_*$, the matching law gives $\ell\approx50$ and predicts, out of sample, a flow at $n_*$ and flows below $n_*$ down to $n_c\approx6.571$. Two numerical methods confirm both predictions down to $n=6.574$, and a computation of the flow as a single orbit gives the end of the branch, $n_c=6.5712$ (appendix~\ref{app:numerics}). These flows are nodeless, positive and stable against static perturbations, and their action lies below that of $\Dz$, since on the plane it equals $-\frac12H(0)$ (section~\ref{sec:endpoint}). Symmetry breaking therefore comes before conformality loss: this R\'enyi defect orders below $n_*$, at $n_c$, while $\Dz$ is still a consistent conformal defect.

The same happens at $\Delta=2.25$ with a stronger attractive quartic, $c_3=-0.7$ and $c_5=\frac1{10}$: the same two methods find nodeless source-free flows at $n=12$ to $14$, below $n_*=14.82$, one of them down to $n\approx11.97$, and the branch ends at $n_c=11.963$. With the attractive quartic alone, a potential unbounded below, there is no positive homogeneous saddle beyond the onset of the hairy phase and no positive source-free solution beyond the threshold of the cutoff ball, so the flow runs away. In the O($N$) model, too, the runaway can set in before the fixed points merge, at an index set by the microscopic coupling \cite{MFS2009}.

\section{Chemical potential, backreaction and finite $\Nn$}
\label{sec:backreaction}

Sections~\ref{sec:renyidefect}--\ref{sec:generic} work in the probe limit, in which the hyperbolic black hole is fixed, at large $\Nn$ and, for the flows, at zero chemical potential. This section follows the merger to imaginary and real chemical potential, adds the backreaction of the matter fields on the metric at first order and returns to the probe limit at finite $\Nn$.

\subsection{Flux: conical monodromy-pinning defects}
\label{sec:flux}

The field equations depend on $\mu$ through $\mu^2$, and on $\Bb_t$ linearly in Maxwell's equation, so the continuation $\mu=ia$ of \eqref{eq:family} is again a solution:
\begin{equation}
eL\rhob=\frac{p\,u^2}{1-a^2u^2/4},\qquad \Bb_\tau=-i\Bb_t=\frac{a\,f}{1-a^2u^2/4},\qquad p^2=\frac12\big(-a^2-\muc^2\big)\Big(\frac{4}{\uh^2}-a^2\Big).
\label{eq:flux}
\end{equation}
It is real and regular for $a^2<-\muc^2(n)$, since $-\muc^2<4/\uh^2$ at every temperature, and therefore requires $n>1$. The Euclidean gauge field has the holonomy $e^{i\alpha}$ around the thermal circle, with $\alpha=2\pi naR$. In the frame without a background field the condensate winds, $\langle O\rangle=a_O\,e^{-ia\tau}/r^2$, with monodromy $e^{-i\alpha}$ around the full cone. The holonomy is that of the twist defect with a U(1) flux, which computes the generating function $Z_n(\alpha)={\rm Tr}\,\rho_A^n e^{i\alpha Q_A}$ of the symmetry-resolved R\'enyi entropies \cite{BHMMMS2013,GoldsteinSela2018}, studied holographically in AdS$_3$ in \cite{ZNM2021}.

The defect breaks the U(1) and the transverse rotations, but preserves a combination of the two. This is the structure of the monodromy-pinning defects constructed in the critical O($2N$) model at large $N$ and in the $\varepsilon$-expansion as infrared fixed points of a monodromy defect deformed by a relevant operator of nonzero transverse spin \cite{KravchukRadcliffe2025}. Here the defect is conical as well, the symmetry is broken by a condensate and the solution is in closed form. Supersymmetric conical monodromy defects with charged mass deformations that wind around the defect were constructed in supergravity in \cite{AGJRR2024}. At $n=1$ the hairy solution does not exist: the pure monodromy defect has no relevant charged operator in the channel $k=0$, of transverse spin $-aR$, since an imaginary chemical potential raises the effective mass of the scalar.

The condensate vanishes on the curve where the charged operator of the normal conical monodromy defect with $k=0$ is marginal, $\Dh=2$:
\begin{equation}
\alpha_c(n)^2=\pi^2\Big[16n^2-\big(1+\sqrt{1+8n^2}\big)^2\Big]=\frac{32\pi^2}{3}(n-1)+O\big((n-1)^2\big).
\label{eq:alphac}
\end{equation}
It reaches $\alpha_c=\pi$ at $n=1.0881$, so for larger $n$ the hairy saddle exists for every $\alpha\in(-\pi,\pi]$. In the hairy phase $b$ changes by $\Delta b=-\frac32\Nn\,nR^4(\muc^2+a^2)^2$, the stiffness is $\rho_s=(L/e^2)\,p^2/(2/R^2-a^2)$, and the separation of the grand potential into a part that depends only on $\mu$ and a part that depends only on $\uh$ \cite{MainPaper} gives
\begin{equation}
\Om_\SF(ia)-\Om_\SF(0)=\frac{L}{e^2}\,a^2\Big(\frac{1}{R^2}-\frac{a^2}{8}\Big),
\end{equation}
independent of $n$. The universal part of the flux dependence is therefore
\begin{equation}
\ln\frac{Z_n(\alpha)}{Z_n(0)}\Big|_{\ln(2R/\delta)}=\frac{\Nn}{\pi^2}\,\frac{\alpha^2}{n}\Big(1-\frac{\alpha^2}{32\pi^2n^2}\Big)\quad\text{(hairy)},\qquad \frac{\Nn}{\pi^2}\,\frac{\alpha^2}{n}\,\frac{R^2}{\uh^2}\quad\text{(normal)}.
\label{eq:fluxgen}
\end{equation}
For the pinned defect, this continues the charged R\'enyi entropies of \cite{MainPaper} across the phase boundary. The flux also moves the merger. In the normal phase with flux, the charged operator with $k=0$ meets its alternate-quantization partner, $\eta=1$, on a line $n_*(\alpha)$ that rises from $n_*$ at $\alpha=0$ to $n_*(\pi)=6.70$, while those with $k=1$, of transverse spin $1/n-aR$, keep real dimensions. Beyond the line the charged operator of the R\'enyi defect with flux has a complex dimension. Unlike at real $\mu$, the Euclidean action is convex in the gauge field here, so the gradient-flow and minimizer arguments of section~\ref{sec:endpoint} apply to the flows of this defect.

We computed these flows at $\alpha=\pi/2$ and $\pi$ for $7.5\le n\le30$, solving for the scalar and $\Bb_\tau$ together, with two numerical methods (appendix~\ref{app:numerics}). They are nodeless and stable against static perturbations, and they end at the winding saddle \eqref{eq:flux}. The flux acts mainly through $\nu$: the walls follow the $\pi/\nu$ law, and $c_0$ is larger at $\alpha=\pi$ than at $\alpha=0$ by only $0.012$. On the sphere the coefficient of $\ln(2R/\delta)$ is that of the winding saddle, so beyond the line the hairy form of \eqref{eq:fluxgen} holds without a source. No source-free flow is found at the points tested below the line (appendix~\ref{app:numerics}). The uniqueness argument of section~\ref{sec:endpoint} and the energy argument of section~\ref{sec:sourcefree} do not extend as stated, since the screening of $\Bb_\tau$ by the condensate adds an attractive term of order $a^2$. Uniqueness and the absence of flows below the line are therefore supported only numerically, although on every flow computed the term, integrated over the flow, is at most $0.22\%$ of the quartic repulsion of the potential.

\subsection{The merger line at real chemical potential}
\label{sec:realmu}

\begin{figure}[t]
\centering
\includegraphics[width=0.62\textwidth]{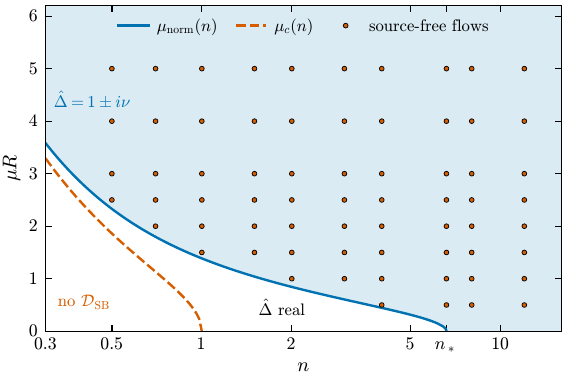}
\caption{The merger line at real chemical potential. In the shaded region above $\mu_{\rm norm}(n)$, which ends at $n_*$ at $\mu=0$, the charged operator of $\Dz$ at transverse spin zero has $\Dh=1\pm i\nu$; below it $\Dh$ is real. The hairy saddle $\Dsb$ exists above $\muc(n)$, which vanishes at $n=1$. Dots: the computed source-free flows, all of them nodeless.}
\label{fig:realmu}
\end{figure}

At real $\mu$ the merger moves to lower index. In this subsection $R=1$. The charged operator of $\Dz$ at transverse spin zero has $\eta=\eta_0(\mu)$, the largest eigenvalue of the radial operator of the normal phase, which increases with $\mu$ at fixed $n$. First-order perturbation theory about the exact zero mode $u^2/(1+\muc^2u^2/4)$ at $\mu^2=\muc^2$, the onset of \eqref{eq:family}, which for $n>1$ lies at imaginary $\mu$, gives $\eta_0=\frac12(\mu^2-\muc^2)+O\big((\mu^2-\muc^2)^2\big)$ at every $n$. At $\mu=0$ the first term is $-M^2$ of eq.~\eqref{eq:project}, since $M^2=\muc^2/2$ by \eqref{eq:mucn}. The eigenvalue $\eta_0$ reaches $1$, where $\Dh_\pm$ meet at $\Dh=1$, on the line $\mu=\mu_{\rm norm}(n)$, the onset of the normalizable instability at real chemical potential \cite{MainPaper}. The line ends at $n_*$ at $\mu=0$, and it reaches $n=1$ at $\mu_{\rm norm}=1.387$ and lower indices beyond (figure~\ref{fig:realmu} and table~\ref{tab:onsets}), so at real chemical potential the merger is not confined to large $n$. With the flux of section~\ref{sec:flux} the line continues to larger $n$, up to $n_*(\pi)=6.70$, so it passes through $\mu^2=0$ at $n_*$. The hairy saddle \eqref{eq:family} exists for $\mu>\muc(n)$, which for $n>1$ is every real $\mu$.

\paragraph{What the merger means here.} At real $\mu$ the R\'enyi defect is not reflection positive along the defect. In the frame without a background field, $\langle O\rangle=a_O\,e^{-\mu\tau}/r^2$ and $\langle O^\dagger\rangle=a_O\,e^{\mu\tau}/r^2$. Reflection positivity in $\tau$ requires $\langle O^\dagger(-\tau)\rangle=\langle O(\tau)\rangle^*$, which holds, while along the defect it requires $\langle O^\dagger(\tau)\rangle=\langle O(\tau)\rangle^*$, which holds only at imaginary $\mu$: a real chemical potential inserts the positive operator $e^{\mu\hat Q/T}$, with $\hat Q$ the charge, rather than a unitary twist. Complex dimensions are therefore allowed, and they occur. In the normal phase the gauge field is $B_{\rm N}=\mu(1-u^2/\uh^2)$, and a charged operator with $k\neq0$ has the transverse spin $k/n+i\mu$: its radial operator is complex, and $\eta$ is complex once $\mu\neq0$, with $|\mathrm{Im}\,\eta|=2|k|\mu$ exactly at $n=1$, where $B_{\rm N}/f=\mu$. A complex dimension alone therefore does not make $\Dz$ inconsistent here. With $k=0$, however, the radial operator depends on $\mu$ only through $B_{\rm N}^2$, so $\eta_0$ is real, and at $\mu_{\rm norm}$ two real fixed points, the two quantizations of $\Dz$, meet at $\Dh=1$ and annihilate, as at $n_*$ (section~\ref{sec:spectrum}). What the merger implies for the flows follows from the action instead, through the two results below.

\paragraph{No source-free flow below the line.} With $P=u^2\chi(u,s)$ and $B_t=B(u,s)$ as in section~\ref{sec:cutoff}, static flows obey
\begin{equation}
\mathcal{L}_u\chi+u\nabla^2\chi+\frac{u}{f}B^2\chi-\frac12u^3\chi^3=0,\qquad fu\,\partial_u\Big(\frac{\partial_uB}{u}\Big)+\nabla^2B-u^2\chi^2B=0,
\label{eq:flowmu}
\end{equation}
with $\mathcal{L}_u$ the operator of section~\ref{sec:cutoff}, which does not depend on $\mu$, and $\nabla^2$ the Laplacian of $H^3$, equal to $\partial_s^2-2\partial_s$ on functions of $s$. The boundary conditions are $B=\mu$ at the AdS boundary, $B=0$ at the horizon and $B=B_{\rm N}$ at the cutoff, where $\chi=0$. Multiplying the scalar equation by $\chi$ and integrating over $u$ and the cutoff region of $H^3$, as in section~\ref{sec:sourcefree}, gives
\begin{equation}
Q_\mu[\chi]+\frac12\int u^3\chi^4=\int\frac{u}{f}\big(B^2-B_{\rm N}^2\big)\chi^2,
\label{eq:noflow}
\end{equation}
where $Q_\mu[\chi]=\int\big[-\chi\mathcal{L}_u\chi+u|\nabla\chi|^2-(u/f)B_{\rm N}^2\chi^2\big]$ is the quadratic form of $\Dz$ at $\mu$, nonnegative for $\mu\le\mu_{\rm norm}$ since $\eta_0\le1$ there. The condensate screens the gauge field, $0\le B\le B_{\rm N}$ by the maximum principle, so the right side is not positive, and $\chi=0$: below the line no source-free flow exists. With flux the same term has the opposite sign, which is why section~\ref{sec:flux} could not exclude such flows.

\paragraph{A Lyapunov function.} At $\mu=0$ the flows are gradient-like (section~\ref{sec:endpoint}). At real $\mu$ the function $H$ of that section, with the gauge field included,
\begin{equation*}
H(s)=\int du\,\Big[\frac12u\chi_s^2+\frac12\chi\mathcal{L}_u\chi-\frac18u^3\chi^4+\frac{u}{2f}B^2\chi^2+\frac{B_u^2}{2u}-\frac{B_s^2}{2uf}\Big],
\end{equation*}
obeys $dH/ds=2\int du\,\big(u\chi_s^2-B_s^2/uf\big)$, which has no sign. A different function decreases. Let $\bar\chi$ and $\Bb_t$ be the homogeneous saddle \eqref{eq:family}, $\delta B=B-\Bb_t$, and
\begin{equation}
\mathcal{K}(s)=\int du\,\Big[u\Big(\frac{\chi^2}{2}-\bar\chi^2\ln\chi\Big)+\frac{(\delta B)^2}{uf}\Big].
\label{eq:Lyap}
\end{equation}
On a positive solution of \eqref{eq:flowmu}, $\mathcal{K}''-2\mathcal{K}'=\mathcal{P}$, with
\begin{equation}
\begin{aligned}
\mathcal{P}=\int du\,\Big\{&uf\Big[\Big(\chi_u-\frac{\chi\bar\chi_u}{\bar\chi}\Big)^2+\Big(\bar\chi_u-\frac{\bar\chi\chi_u}{\chi}\Big)^2\Big]+u\chi_s^2\Big(1+\frac{\bar\chi^2}{\chi^2}\Big)+\frac12u^3\big(\chi^2-\bar\chi^2\big)^2\\
&+\frac{2(\partial_u\delta B)^2}{u}+\frac{2(\partial_s\delta B)^2}{uf}+\frac{u}{f}\big(\chi^2+\bar\chi^2\big)(\delta B)^2\Big\}\ge0.
\end{aligned}
\label{eq:Pos}
\end{equation}
To derive it, replace $\nabla^2\chi$ and $\nabla^2\delta B$ in $\nabla^2\mathcal{K}$ by eqs.~\eqref{eq:flowmu}, add the scalar and Maxwell equations of the saddle times $(\chi^2-\bar\chi^2)/\bar\chi$ and $2\,\delta B/uf$, respectively, and integrate by parts in $u$, with $\delta B=0$ at the AdS boundary and the horizon. The gradient terms combine into squares as in Picone's identity (appendix~\ref{app:proof}), and the gauge terms through $-(\chi^2-\bar\chi^2)(B^2-\Bb_t^2)+2\,\delta B\,(\chi^2B-\bar\chi^2\Bb_t)=(\chi^2+\bar\chi^2)(\delta B)^2$. For a bounded flow, integrating $(e^{-2s}\mathcal{K}')'=e^{-2s}\mathcal{P}$ from $s$ to infinity gives
\begin{equation}
\mathcal{K}'(s)=-\int_s^\infty dt\,e^{-2(t-s)}\,\mathcal{P}(t)<0.
\label{eq:Gprime}
\end{equation}
So $\mathcal{K}$ decreases along every bounded positive flow, which therefore cannot be periodic, and since $\mathcal{K}$ is bounded below, $\mathcal{P}$ is integrable along the flow, which approaches $\Dsb$, the only positive configuration with $\mathcal{P}=0$. The positivity of $\mathcal{P}$ needs a real gauge field, $\mu^2>0$; with flux it fails, and there $H$ is monotonic instead. For a potential other than the quartic, $\frac12u^3(\chi^2-\bar\chi^2)^2$ becomes a positive multiple of $(\chi^2-\bar\chi^2)\big[V'(\rho)/\rho-V'(\rhob)/\rhob\big]$, nonnegative when $V'(\rho)/\rho$ increases, the condition of the uniqueness argument of section~\ref{sec:endpoint}.

\paragraph{Computed flows.} We solved eqs.~\eqref{eq:flowmu} for $67$ source-free flows with $0.5\le n\le12$ and $0.5\le\mu\le5$, all beyond the line (figure~\ref{fig:realmu}). On the plane the far end is held at $\Dsb$, as in section~\ref{sec:cutoff}, and the walls do not depend on $S$ (appendix~\ref{app:numerics}). All are nodeless. $H'$ changes sign on three of the five flows on which we evaluated it. Near the line the walls follow the $\pi/\nu$ law of section~\ref{sec:bkt}, with $\nu=\sqrt{\eta_0-1}$: the coefficient of $1/\nu$ is $\pi$ at $n=1$ and $n=3$. Below the line, with a pinning field at the cutoff, the wall moves at the rate $1/(2-\Dh_+)$ of section~\ref{sec:pinning}. At larger $\mu$ the two leading irrelevant operators of $\Dsb$ at transverse spin zero can form a complex-conjugate pair, as at $(n,\mu)=(1,3)$, where $\Dh-2=2.333\mp0.436\,i$. There $\Dsb$ remains attractive, and the flows approach it in a damped spiral, which by eq.~\eqref{eq:Gprime} cannot close into a cycle. We do not treat this spectrum further.

\paragraph{The sphere.} We expect the proposition of section~\ref{sec:sphere} to hold at real $\mu$, with $\nu=\sqrt{\eta_0(\mu)-1}$. The Euclidean action is concave in $B$, so eliminating $B$ leaves a reduced action of $\chi$ that is real, strictly convex in $\chi^2$ and bounded below, with the quadratic form of $\Dz$ at small $\chi$: the setting of the argument of Brezis and Oswald used in appendix~\ref{app:proof}. We have not written out its steps with the nonlocal term that the elimination of $B$ leaves. The computed solutions agree with it: on the ball, with nothing imposed in the interior, they switch on at the threshold $\nu=\pi/\zeta_{\max}$ and end at $\Dsb$ in the interior, and the coefficient of $\ln(2R/\delta)$ in their free energy is the homogeneous one, $-2\pi\Om_{\rm hom}$, with $\Om_{\rm hom}=-(L/8e^2)(\mu^2-\muc^2)^2$ (appendix~\ref{app:numerics}). For these solutions the universal term beyond the line is therefore that of $\Dsb$, as beyond $n_*$ at $\mu=0$. At $n=1$, where the line lies at $\mu=1.387$, the same holds for the entanglement entropy at real chemical potential.

\subsection{Backreaction at first order}
\label{sec:brmu0}

The action of \cite{MainPaper} contains the Einstein--Hilbert term $(1/2\kappa_5^2)\int d^5x\sqrt{-g}\,(\mathcal{R}+12/L^2)$, and the backreaction parameter is
\begin{equation}
\ve\equiv\frac{\kappa_5^2}{e^2L^2}.
\label{eq:kb}
\end{equation}
It is of order $C_J/C_T$ \cite{MainPaper}, and the probe limit is $\ve\to0$ at fixed $\Nn$. On the hairy saddle \eqref{eq:family} the matter stress tensor corrects the metric at relative order $\ve$, by a correction rational in $u$, and the grand potential at first order needs only this metric: the first-order corrections of the matter fields drop out by stationarity. Appendix~\ref{app:backreaction} gives the method and the onsets at real chemical potential. We write the results below in terms of the variable $v\equiv\uh^2/R^2$, which runs from $1$ at $n=1$ to $2$ at large $n$.

\paragraph{The merger.} The merger itself cannot move. At $\mu=0$ the normal phase has $\Psi=A=0$ and therefore does not source the metric, and since the scalar enters the Einstein equations only quadratically, its linearized equation on that background contains no $\ve$. Hence $n_*$ and the spectrum of $\Dz$ at $\mu=0$ are the same at every order in $\ve$, classically in the bulk. This is also why the onset of \cite{BMM2013}, obtained with dynamical gravity from the linear scalar on the uncharged black hole, equals the value in the probe limit.

\paragraph{The hairy saddle.} The hairy saddle does source the metric. At $\mu=0$, eq.~\eqref{eq:Om2int} gives
\begin{equation}
\Om_\SF-\Om_\N=-\frac{L}{8e^2}\,\muc^4\big[1+\ve\,W(v)\big],\qquad
W(v)=-\frac{4\,(6+7v-v^2)}{9\,(2+v)},
\label{eq:renyi1}
\end{equation}
with $\muc$ the probe value \eqref{eq:mucn}. The ratio $W$ equals $-\frac{16}{9}$ at both ends, $n\to1$ and $n\to\infty$, and varies little in between. Since the hairy saddle does not exist at $T_0$, its contribution to the R\'enyi entropies is multiplied by the same factor, and so is $\Delta b$, eq.~\eqref{eq:db}: backreaction reduces both at every index, by a factor between $1-1.81\,\ve$ and $1-\frac{16}{9}\ve$. Since eq.~\eqref{eq:dh} is linear in $\Om_\SF-\Om_\N$, the conformal weight at first order becomes
\begin{equation}
\Delta h_n=\frac{n\Nn}{12\pi}\,\frac{d}{dn}\Big(n\,\muc^4R^4\big[1+\ve\,W(v)\big]\Big).
\label{eq:dh1}
\end{equation}

\paragraph{The flows.} Below $n_*$, eqs.~\eqref{eq:renyi1} and \eqref{eq:dh1} are the data of the pinned defect $\Dsb$ (section~\ref{sec:spectrum}). Beyond $n_*$ they give the first-order corrections to the universal R\'enyi term of section~\ref{sec:sphere} and to the jumps of $b$ and $h_n$ at $n_*$, if the source-free flow still ends at $\Dsb$ at first order. Away from $n_*$ we argue that it does. On the cutoff ball the probe flow is a nondegenerate minimizer (the proposition of section~\ref{sec:sphere}), so we expect the implicit function theorem, applied to the coupled Einstein--scalar problem, to continue it to small $\ve$. This needs boundary conditions for the metric at the cutoff, which we do not specify, and no zero mode of its static perturbations, which we have not shown. By the layer argument of section~\ref{sec:endpoint}, the universal term of the continued flow is then that of the backreacted homogeneous saddle. Near $n_*$ the lowest eigenvalue of the stability operator goes to zero, and the expansion is not uniform; we expect that only strong backreaction could change the outcome there, by moving the zero of the tail of section~\ref{sec:pinu} far toward the ultraviolet. At first order the amplitude operators of $\Dsb$ at $\mu=0$, with transverse spin $0$ and $1/n$, move by at most about $0.05\,\ve$ at the indices computed, up to $n=12$, and stay irrelevant.

\subsection{Finite $\Nn$}
\label{sec:finiteN}

At infinite $\Nn$, $\Dsb$ breaks the U(1) with a classical condensate. At finite $\Nn$ this cannot persist in the strict sense. Cuomo and Zhang showed that a continuous internal symmetry cannot be broken spontaneously on a surface defect of a conformal field theory if the defect flow ends at a scale-invariant fixed point \cite{CuomoZhang2023}. They also showed that the generic weakly coupled alternative is a defect conformal field theory coupled, through the tilt, to a two-dimensional sigma model of the Goldstone modes, with a marginally irrelevant coupling; for boundaries of the O($N$) model this is the extraordinary-log class of \cite{Metlitski2020}. For a U(1), the beta function of the sigma-model coupling $g$, with stiffness $1/g^2$, has the coefficient $\alpha_{\rm CZ}=\frac{\pi}{2}C_{\tilt}>0$ \cite{CuomoZhang2023}. The coupling therefore flows to zero logarithmically, and the order parameter along the defect decays as
\begin{equation}
\langle O(\vec y)\,O^\dagger(0)\rangle\sim(\ln|\vec y|)^{-q},\qquad q=\frac{1}{2\pi\alpha_{\rm CZ}}=\frac{1}{\pi^2C_{\tilt}}.
\label{eq:q}
\end{equation}
With the tilt normalization \eqref{eq:Ct}, $q=C_O/(2n\,a_O^2)$ at $\mu=0$, which is $3/\big(32\Nn(n-1)\big)$ near $n=1$. The exponent is of order $1/\Nn$, with $q\Nn=0.038$ at $n_*$, falling as $1/4n$ at large $n$, so the decay is very slow, and vortices of the orientation on the defect, whose action grows with $C_{\tilt}$, are suppressed when $\Nn$ is large.

The logarithmic class is the fate of a \emph{spontaneous} $\Dsb$. Below $n_*$, $\Dsb$ is reached only with a pinning field, which also fixes the orientation in the infrared, and the order parameter does not decay. Beyond $n_*$ the R\'enyi defect itself flows to $\Dsb$ (section~\ref{sec:flow}) with no source, and at large but finite $\Nn$ the correlation function \eqref{eq:q} then describes the R\'enyi defect beyond the scale $z_*$. This combines the theorem and the effective theory of \cite{CuomoZhang2023} with the large-$\Nn$ value of $C_{\tilt}$; the corrections of order $1/\Nn$ themselves, from loops in the bulk, are not computed here.

A real scalar with a $\mathbb{Z}_2$ symmetry gives the same flows at $\mu=0$, where the phase of $\Psi$ is constant. Its fate at finite $\Nn$ can differ: the theorem of \cite{CuomoZhang2023} concerns continuous symmetries, and a discrete symmetry can break spontaneously in two dimensions, as on the boundary of the three-dimensional Ising model at the extraordinary transition \cite{BD1994}. With no tilt operator there is no logarithmic decay, and nothing forbids the ordered defect at finite $\Nn$.

\section{Discussion}
\label{sec:discussion}

\paragraph{Lattice and large-$N$ R\'enyi defects.} Quantum Monte Carlo simulations of R\'enyi defects at $(2+1)$-dimensional O(3) critical points find several defect universality classes, selected by the microscopic entanglement cut, defect dimensions of the order parameter that decrease with $n$ and, for one cut, a transition to an ordered regime at $n\approx3$--$4$ \cite{ZWCY2026}. The microscopic symmetry of the cut also matters for the second R\'enyi entropy at a deconfined critical point \cite{WLY2026}. The setting differs from ours in essential ways: the defects are lines, the symmetry is O(3) and the measured dimensions already lie below $\frac12$, the defect BF bound of a line, at $n=2$, where a merger criterion valid at large $\Nn$ does not apply. What carries over is qualitative: the dimension of the charged defect operator decreases with $n$, and a R\'enyi defect can order at large $n$. In holography the index at which it orders can be small, below $n=2$ for $\Delta<1.438$ (section~\ref{sec:crossing}). In field theory, with normalizable modes on $H^2$, the R\'enyi defect of the large-$N$ O($N$) model has a symmetry-breaking instability that sets in at $n=1$ \cite{HLS2018}, where the fundamental field, of dimension $\frac12$, sits at the defect BF bound of a line. For $n>1$, with the alternate quantization allowed, the gap equation has a unique real solution, found numerically in \cite{MFS2009}, of defect dimension $\frac12-\frac14(n-1)$ near $n=1$, the slope of section~\ref{sec:marginal}. On this solution the two quantizations, $\frac12\mp\frac14(n-1)$ to first order, meet at $n=1$ and stay real, unlike those of section~\ref{sec:spectrum}.

\paragraph{Relation to earlier work.} The two instabilities and the onset $n_*$ at $\Delta=2$ are those of \cite{BMM2013}; the onset at exactly $n=1$ at the BF bound was found in \cite{DMRS2010}, and the mass at which the constant mode is a zero mode at $T_0$ in general dimension in \cite{BelinMaloney2014}. The impurity operator of \cite{BHMM2014} drew on the O($N$) results of \cite{MFS2009} and on a transition in the R\'enyi--Shannon entropy when a boundary operator becomes relevant \cite{SMP2011}; the endpoint constructed here is a defect of that kind, at zero and at real chemical potential. Mergers and annihilations of defect fixed points also occur for spin impurities, with runaway regimes \cite{CKMRimp2022} and a fixed-point annihilation \cite{Nahum2022}, for defects in scalar field theories in a semiclassical limit \cite{RGR2022} and at the boundary of the O($N$) model \cite{KM2023}. The flows, phases and Weyl anomalies of surface defects in the free and critical O($N$) models were studied in \cite{RMZ2023,Trepanier2023,GiombiLiu2023}, and a localized magnetic field, the pinning field of a line, in \cite{CKM2022}. In probe-brane holography, defects show BKT-type transitions out of a conformal phase \cite{JKST2010,KLP2011,GKS2012}, and in some embeddings of the last the defect also breaks chiral symmetry. The holographic twist operator was treated as a defect CFT in \cite{BDF2016}. The picture of a symmetry-breaking defect as a condensate on the AdS factor of the conformal frame, the absence of spontaneous breaking on surface defects and the logarithmic class are from \cite{CuomoZhang2023}, with the boundary analog in \cite{Metlitski2020}, and the tilt and displacement operators of such defects were studied in \cite{BdSM2026,BDS2026}. Related work computed one-point functions at holographic monodromy defects \cite{CCR2026}, the entanglement entropy of spheres centered on superconformal monodromy defects, with or without a conical singularity \cite{CLRR2025}, and nonsupersymmetric holographic flows of surface defects \cite{PRRZ2026}, and described vortices of holographic superfluids as conformal defects \cite{DHIS2013}. We have not found an earlier reading of the homogeneous hairy saddle as a conformal defect, or a construction of the endpoint of the R\'enyi defect beyond its normalizable instability.

\paragraph{Scope.} The flows are computed in the probe limit and at leading order in $1/\Nn$: at zero chemical potential for $6.59\le n\le30$ in the model of section~\ref{sec:renyidefect} and, each over its own range of $n$, at three further scalar masses, with a fourth for existence only (section~\ref{sec:generic}); with flux for $7.5\le n\le30$; and at real chemical potential for $0.5\le n\le12$ (section~\ref{sec:realmu}). The universal term is taken with $R/\delta\to\infty$ first; at finite $R/\delta$ the R\'enyi entropy follows $\Dz$ up to the threshold of the cutoff ball (section~\ref{sec:sphere}). Not covered are the structure of the complex spectrum of $\Dsb$ at larger real chemical potential, backreaction beyond first order and corrections in $1/\Nn$ beyond the effective theory of section~\ref{sec:finiteN}. Other bulk fields and $d\neq4$, including the setting of \cite{BHMM2014}, are not covered either, nor are cutoff prescriptions other than the two of section~\ref{sec:bkt}, $\chi=0$ and the Robin condition with $\Gamma>1$, which differ in $c_0$ but not in the coefficient of $1/\nu$ or the universal term. Stability of the flows in real time is shown for linear perturbations, in the probe limit at $\mu=0$ (section~\ref{sec:sourcefree}), but not nonlinearly or with backreaction.

\paragraph{Top-down theories.} The model of this paper is bottom-up. In maximally supersymmetric Yang--Mills theory at large $N$ and strong coupling, the scalars of the $\mathbf{20}'$ have $\Delta=2$ \cite{BelinMaloney2014} and obey the same linear equation on the uncharged hyperbolic black hole, so the merger is at the same $n_*$. The homogeneous saddles of this theory include some of the analytic hairy black holes of \cite{Ren2019}, which lie in a truncation of the dual supergravity. The backreaction is of order one there, however, and the potential of these scalars has a cubic term and is unbounded below, so the conditions of the argument of section~\ref{sec:generic} do not hold. Nor are these black holes the endpoint: on them the $\mathbf{20}'$ operators that transform as a traceless symmetric tensor of the unbroken SO(4) have complex defect dimensions for every $n>1.38$, since the largest $\eta$ of their radial operator increases with $n$. Whether the R\'enyi defect of this theory flows beyond $n_*$ to another homogeneous saddle, orders before $n_*$ or runs away is open.

\paragraph{Open questions.}
\begin{itemize}
\item For $3.28<n<31.1$ the R\'enyi defect has relevant charged operators of transverse spin $1/n$, and for $6.08<n<59.2$ of spin $2/n$. Deforming by them should give spinning conical defects, the analogs of \cite{KravchukRadcliffe2025}, which are not in the closed-form family and are not constructed here.
\item With flux, the uniqueness of the source-free flow and the absence of flows below the merger line are supported only numerically (section~\ref{sec:flux}); static stability in the channels of nonzero transverse spin was not checked there.
\item At real chemical potential the proposition of section~\ref{sec:sphere} is expected but not proven: the steps of appendix~\ref{app:proof} with the nonlocal term of the reduced action are not written out (section~\ref{sec:realmu}). The Lyapunov function covers static flows; flows that depend on $\tau$ are not excluded from cycling.
\item The spectrum of $\Dsb$ in the phase and vector sectors beyond the tilt has not been computed, nor has the displacement operator, whose normalization fixes the remaining anomaly coefficient $d_1$ of $\Dsb$ \cite{JORR2018}.
\item The corrections of order $1/\Nn$, which by section~\ref{sec:finiteN} turn the classical breaking into a logarithmic decay, require the fluctuation determinant of the flow solution.
\item With backreaction, only the first-order metric and the grand potential of the hairy phase are known in closed form; its first-order matter fields, which the one-point coefficient and the tilt normalization need, were not computed. Whether strong backreaction can produce a source-free flow already at $n_*$, and so move the transition below it, is open.
\end{itemize}

\section*{Acknowledgments}
Claude Opus 5.5 (Anthropic) assisted with the scientific reasoning, the derivations, the literature search and the drafting of the text. The author directed it by choosing the questions and deciding what to pursue and include. The author rederived the analytic results, checked the arguments, reviewed the numerical methods and their controls, and is solely responsible for the content, including any errors. The programs used in this work are available from the author upon~reasonable~request.

\appendix
\section{Source-free solutions on the cutoff ball}
\label{app:proof}

This appendix proves the proposition of section~\ref{sec:sphere}: parts (i)--(iv) in the paragraphs of those names, part (v) in those after them and, in the last paragraph, that the coefficient of $\ln(2R/\delta)$ in the free energy is the homogeneous one. We set $R=1$, so that $f=(1-u^2/\uh^2)\big(1-(\uh^2-1)u^2/\uh^2\big)$ and $p^2=8(\uh^2-1)/\uh^4$, with $1<\uh^2<2$. Let $B$ be the cutoff ball and $\lambda_1=1+\pi^2/\zeta_{\max}^2$ the lowest Dirichlet eigenvalue of $-\nabla^2$ on it.

\paragraph{The sphere variable.} Put $u=\uh\cos(\xi/2)$, with $\xi\in[0,\pi]$, and let $\tilde f=f/(1-u^2/\uh^2)$. It equals $1-(\uh^2-1)\cos^2(\xi/2)$ and lies between $2-\uh^2>0$ and $1$. Then
\begin{equation}
uf\,\partial_u\chi=-\tilde f\sin\xi\,\partial_\xi\chi,\qquad u\,du=-\tfrac14\uh^2\sin\xi\,d\xi.
\end{equation}
Read $\xi$ as the polar angle of a unit two-sphere $S^2$, and let $\mathcal{M}=S^2\times B$. Eq.~\eqref{eq:flowpde} on the ball, divided by $u$, becomes
\begin{equation}
\nabla\cdot(\mathcal{G}\nabla\chi)+\tfrac12u^2\big(p^2-\chi^2\big)\chi=0,
\label{eq:sphereform}
\end{equation}
where $\mathcal{G}$ is $4\uh^{-2}\tilde f$ on $S^2$ and $1$ on $B$, and $u^2=\frac12\uh^2(1+\cos\xi)$. The free energy \eqref{eq:Eball} becomes $E=\frac{\uh^2}{8\pi}\int_{\mathcal{M}}\big[|\nabla\chi|_{\mathcal{G}}^2+\frac14u^2(\chi^2-p^2)^2-\frac14u^2p^4\big]$. The coefficients are smooth on $\mathcal{M}$, and $\mathcal{G}$ is bounded between positive constants, so the operator is uniformly elliptic on a compact manifold whose only boundary, the cutoff $S^2\times\partial B$, is smooth. The AdS boundary and the horizon are the poles $\xi=\pi$ and $\xi=0$, inside $\mathcal{M}$, and the weighted norm $\int u\big(f\chi_u^2+|\nabla\chi|^2+\chi^2\big)$ that $E$ controls is the norm of $H^1(\mathcal{M})$.

\paragraph{Admissible solutions.} Near $u=0$ a solution is a constant plus a multiple of $\ln u$, the source, and near the horizon a constant plus a multiple of $\ln(\uh-u)$. In the sphere variable these logarithms are $\ln(\pi-\xi)$ and $2\ln\xi$ up to constants, the Green's function of the two-sphere at a pole, so the solutions of the proposition, which have neither, are the bounded ones. A bounded solution of eq.~\eqref{eq:sphereform} away from the poles is a weak solution on all of $\mathcal{M}$, since the poles times $B$ have codimension two and zero capacity. A function that vanishes within $r_0^2$ of the poles, equals one beyond $r_0$ and is linear in the logarithm of the distance in between has a gradient whose $L^2$ norm vanishes as $r_0\to0$. Testing the equation with it shows that the solution lies in $H^1(\mathcal{M})$ and solves the equation weakly across the poles. Elliptic regularity \cite[chs.~8 and~9]{GilbargTrudinger2001} then makes it smooth inside $\mathcal{M}$, poles included, and $C^{1,\beta}$ up to the cutoff, and $|\chi|\le p$ follows by testing the equation with $\max(\chi-p,0)$ and $\max(-\chi-p,0)$.

\paragraph{Part (i).} With $t=\cos\xi$, the Rayleigh quotient of the radial operator for a trial function $\phi(t)$ is $4\big[(\uh^2-1)X_\phi-Y_\phi\big]/(\uh^2Z_\phi)$, with $X_\phi=\frac12\int(1+t)\big[\phi^2+(1-t^2)\phi_t^2\big]$, $Y_\phi=\int(1-t^2)\phi_t^2$ and $Z_\phi=\int\phi^2$ over $[-1,1]$, which do not depend on $n$. Its derivative with respect to $\uh^2$ is $4(X_\phi+Y_\phi)/(\uh^4Z_\phi)>0$. The radial operator has a compact resolvent on $S^2$, so its largest eigenvalue $\eta_0$ is attained, and since $\uh$ increases with $n$, $\eta_0(n)$ increases strictly. Hence $\eta_0>1$ exactly for $n>n_*$, and $\eta_0<\lambda_1$ for every $n\le n_*$. Multiplying eq.~\eqref{eq:sphereform} by $\chi$ and integrating over $\mathcal{M}$, with no boundary term since $\chi$ vanishes at the cutoff, gives $Q[\chi]+\frac12\int_{\mathcal{M}}u^2\chi^4=0$, with $Q[\chi]=\int_{\mathcal{M}}\big[|\nabla\chi|_{\mathcal{G}}^2-\frac12p^2u^2\chi^2\big]\ge(\lambda_1-\eta_0)\int_{\mathcal{M}}\chi^2$: the $S^2$ part is at least $-\eta_0\int\chi^2$ by the definition of $\eta_0$, and the gradient along $B$ gives at least $\lambda_1\int\chi^2$. If $\eta_0\le\lambda_1$ the quartic integral vanishes, so $\chi=0$. On the half-line of section~\ref{sec:sourcefree} the same identity holds for solutions that stay bounded as $s\to\infty$, with $1$ in place of $\lambda_1$, since the boundary term at large $s$ then vanishes.

\paragraph{Part (ii).} The quartic term is nonnegative, so $E$ is bounded below by the gradient term minus a constant, and the Dirichlet bound on $B$ makes $E$ control the norm of $H^1_0(\mathcal{M})$. A minimizing sequence is therefore bounded; by the Rellich--Kondrachov theorem \cite[ch.~7]{GilbargTrudinger2001} a subsequence converges weakly in $H^1$ and strongly in $L^2$, the gradient term is weakly lower semicontinuous and the potential term obeys Fatou's lemma, so $E$ has a minimizer. The product of the nodeless radial mode and the Dirichlet ground state of $B$, $\sin(\pi\zeta/\zeta_{\max})/\sinh\zeta$, has $Q=(\lambda_1-\eta_0)\int\chi^2<0$, so its small multiples have $E<0=E[0]$, and the minimizer is not zero. Replacing it by $\min(|\chi|,p)$ does not raise $E$, so a minimizer with $0\le\chi\le p$ exists. It is a bounded weak solution, hence smooth, and the strong maximum principle \cite[thm.~3.5]{GilbargTrudinger2001}, applied to $\chi$ and to $p-\chi$, gives $0<\chi<p$ inside $\mathcal{M}$, which contains the poles.

\paragraph{Part (iii).} Let $\chi_1$ and $\chi_2$ be positive solutions. Both vanish on the smooth boundary of $\mathcal{M}$, where Hopf's lemma \cite[lemma~3.4]{GilbargTrudinger2001} gives them a nonzero normal derivative, so $\chi_1/\chi_2$ and $\chi_2/\chi_1$ are bounded \cite{BrezisOswald1986}. The functions $(\chi_1^2-\chi_2^2)/\chi_i$ are then admissible test functions, and subtracting the two tested equations gives, by Picone's identity,
\begin{equation}
-\frac12\int_{\mathcal{M}}u^2\big(\chi_1^2-\chi_2^2\big)^2=\int_{\mathcal{M}}\Big[\Big|\nabla\chi_1-\frac{\chi_1}{\chi_2}\nabla\chi_2\Big|_{\mathcal{G}}^2+\Big|\nabla\chi_2-\frac{\chi_2}{\chi_1}\nabla\chi_1\Big|_{\mathcal{G}}^2\Big]\ge0,
\end{equation}
so $\chi_1=\chi_2$. This is the uniqueness proof of \cite{BrezisOswald1986}, with $\nabla\cdot(\mathcal{G}\nabla)$ in place of the Laplacian. In the sphere variable the stability operator of section~\ref{sec:sourcefree} is $J=J_\theta-u^2\chi^2$, with $J_\theta=\nabla\cdot(\mathcal{G}\nabla)+\frac12u^2(p^2-\chi^2)$. Since $J_\theta\chi=0$ and $\chi>0$, $\chi$ is the ground state of $J_\theta$, whose largest eigenvalue is therefore $0$, so $\int\psi J\psi\le-\int u^2\chi^2\psi^2<0$ for $\psi\neq0$, and hence the largest eigenvalue of $J$ is negative.

\paragraph{Part (iv).} By part (ii) the minimizer of $E$ can be taken positive, and by part (iii) it is the positive solution. For a complex scalar and a gauge field at $\mu=0$, the diamagnetic inequality $|D\Psi|\ge\big|\nabla|\Psi|\big|$ \cite[thm.~7.21]{LiebLoss2001} and $F_{ab}F^{ab}\ge0$ bound the action of the probe sector below by its value at $|\Psi|$ with no gauge field, with equality only for a flat gauge field and a covariantly constant phase. If $|\Psi|$ depends on $\tau$, that value is the integral over $\tau$ of $(L/2e^2)E$ plus a nonnegative term in $(\partial_\tau|\Psi|)^2$. Up to a global phase and a gauge transformation, the minimum of the action is therefore the positive solution $\chi$ of part (iii).

\paragraph{The interior limit.} Let $\chi_B$ be the positive solution on a ball $B$, and $\chi_{\zeta_1}(0)$ the value at its center of the solution on a ball of radius $\zeta_1$; by part (iii) and the isometries of $H^3$, $\chi_{\zeta_1}(0)$ does not depend on the center. If $\chi'$ is a bounded positive solution on a region of $H^3$ that contains the closure of $B$, then $\chi_B\le\chi'$ on $B$. Indeed, $\chi'$ is bounded below by a positive constant on $S^2\times\bar B$, so $(\chi_B^2-\chi'^2)^+/\chi_B$ and $(\chi_B^2-\chi'^2)^+/\chi'$ are admissible test functions, and the computation of part (iii), with $(\chi_B^2-\chi'^2)^+$ in place of $\chi_1^2-\chi_2^2$, shows that this function vanishes (compare \cite{BrezisKamin1992}). For concentric balls, $\chi_B$ therefore increases with the radius. It is bounded by $p$, so by elliptic estimates it converges on $S^2\times H^3$, with its derivatives and uniformly on compact sets, to a solution there. The limit does not depend on the center, since a ball around one point contains, and lies inside, balls around another point whose radii differ from its own by slightly more than the distance between the points. It is therefore invariant under the isometries of $H^3$, hence a function on $S^2$ alone, and integrating eq.~\eqref{eq:sphereform} over $S^2$ gives $\int_{S^2}u^2(p^2-\chi^2)\chi=0$ with $0<\chi\le p$. So the limit is $p$, and $\chi_{\zeta_1}(0)\to p$ uniformly on $S^2$. Finally, the ball of radius $\zeta_1$ around a point farther than $\zeta_1$ from the cutoff of $B$ lies inside $B$, so $\chi_B\ge\chi_{\zeta_1}(0)$ at all points at least $\zeta_1$ from the cutoff, for any radius of $B$. In the interior the flow thus ends at $\Dsb$, uniformly in $\zeta_{\max}$.

\paragraph{The rate.} At these points $p-\chi_B$ solves $\nabla\cdot\big(\mathcal{G}\nabla(p-\chi_B)\big)=\frac12u^2(p+\chi_B)\chi_B\,(p-\chi_B)$, with a coefficient at least $U=\frac12u^2\big(p+\chi_{\zeta_1}(0)\big)\chi_{\zeta_1}(0)$. Let $\eta$ be the largest eigenvalue of $4\uh^{-2}\nabla\cdot(\tilde f\nabla)-U$ on $S^2$, which is negative since $U\ge0$ is not zero, and $\Dh=1+\sqrt{1-\eta}$. The product of its positive eigenfunction and $\sinh\big((\Dh-1)\zeta\big)/\sinh\zeta$ solves the equation with $U$ in place of the coefficient, and on the ball of points at least $\zeta_1$ from the cutoff the maximum principle bounds $p-\chi_B$ by a multiple of it. Hence $p-\chi_B\le C\,e^{-(\Dh-2)s}$ at a distance $s$ from the cutoff, with $C$ independent of $\zeta_{\max}$. As $\zeta_1\to\infty$, $U$ tends to $u^2p^2$, for which the operator is the radial operator of the amplitude fluctuations of $\Dsb$ and $\Dh=\Dh_{\rm amp}$. So $p-\chi_B$ decays into the interior at every rate below $\Dh_{\rm amp}-2$, uniformly in $\zeta_{\max}$; only the value of this rate, $0.561$--$0.569$ for $8\le n\le30$, is numerical, as appendix~\ref{app:numerics} shows.

\paragraph{The planar flow.} On the planar region $s>0$ the direct method of part (ii) applies, since the measure $e^{-2s}ds$ is finite. With the nodeless radial mode times $e^s\sin(\pi s/s_1)$ on $0\le s\le s_1$, $s_1>\pi/\nu$, as trial function, it gives a bounded positive minimizer $\chi_{\rm pl}$. Balls inside the region give $\chi_{\rm pl}(s)\ge\chi_{\zeta_1}(0)$ for $s\ge\zeta_1$, so the identity of part (iii), whose boundary term at large $s$ then vanishes, shows that $\chi_{\rm pl}$ is the only bounded positive solution, and the barrier shows that it tends to $p$ as $s\to\infty$ at the same rate. A copy of the planar region whose cutoff touches that of $B$ from outside contains $B$ (in the upper half-space model $B$ is a Euclidean ball and the cutoff a horizontal plane), and moved outward by any positive distance it contains the closure of $B$. At a point at a distance $s$ from the cutoff of $B$, the copy that touches it at the nearest point therefore gives $\chi_B\le\chi_{\rm pl}(s)$, by the comparison above and the continuity of $\chi_{\rm pl}$. Balls whose cutoffs touch that of $B$ from inside lie in $B$, so the same comparison makes the profile of $\chi_B$ in $s$ increase with $\zeta_{\max}$; its limit is a bounded positive solution of the planar problem, hence $\chi_{\rm pl}$ by the uniqueness above. The maximum principle with $u^2p^2$ in place of $U$ then gives $p-\chi_B\ge c\,e^{-(\Dh_{\rm amp}-2)\zeta_{\max}}$ at the center, with $c$ independent of $\zeta_{\max}$, so the rate is exactly $\Dh_{\rm amp}-2$.

\paragraph{The universal term.} Let $\Om_{\rm loc}$ be $L/2e^2$ times the integrand of eq.~\eqref{eq:Eball} integrated over $u$, which on a solution has the same integral as $\Om$. It equals $\Om_{\rm hom}$ for $\chi=p$, and $\Om_{\rm loc}-\Om_{\rm hom}$ is a sum of squares, which by the bounds above decays exponentially with $s$, uniformly in $\zeta_{\max}$. The planar flow moved onto the ball, $\chi_{\rm pl}(\zeta_{\max}-\zeta)$, and the ball solution moved onto the half-line and continued by its central value are admissible trial functions for the two minimization problems. They bound $E$ from above and below, and with $4\pi\sinh^2(\zeta_{\max}-s)=\pi e^{2\zeta_{\max}}e^{-2s}-2\pi+\pi e^{-2\zeta_{\max}}e^{2s}$ the two bounds agree up to terms that vanish as $\zeta_{\max}\to\infty$:
\begin{equation}
\frac{L}{2e^2}\,E=\pi e^{2\zeta_{\max}}\!\int_0^\infty\!ds\,e^{-2s}\,\Om_{\rm loc}^{\rm pl}-2\pi\Om_{\rm hom}\,\zeta_{\max}-2\pi\!\int_0^\infty\!ds\,\big(\Om_{\rm loc}^{\rm pl}-\Om_{\rm hom}\big)+o(1),
\label{eq:universalproof}
\end{equation}
where $\Om_{\rm loc}^{\rm pl}(s)$ is $\Om_{\rm loc}$ on the planar flow; the constant term also uses the convergence of the profile, with the exponential bound for dominated convergence. The coefficient of $\ln(2R/\delta)$ in the free energy is therefore $-2\pi\Om_{\rm hom}$, the homogeneous one; the area term and the constant are set by the planar flow.

\section{Backreaction}
\label{app:backreaction}

\paragraph{The charged background.} At $\mu\neq0$ the normal phase is the charged hyperbolic black hole, on which the gauge field backreacts at order $\ve$, and both onsets move. Since $\Psi=0$ there, its metric is that of the hyperbolic Reissner--Nordstr\"om--AdS$_5$ black hole at every order in $\ve$. In the coordinates of eq.~\eqref{eq:metric}, with $L=R=1$, the solution reads
\begin{equation}
f=1-u^2-m\,u^4+\frac{2\ve\mu^2}{3}\,\frac{u^6}{\uh^4},\qquad \Bb_t=\mu\Big(1-\frac{u^2}{\uh^2}\Big),\qquad \uh=\frac{4n}{1+\sqrt{1+8n^2\big(1+2\ve\mu^2/3\big)}},
\label{eq:RN}
\end{equation}
at $T=T_0/n$, with $m$ fixed by $f(\uh)=0$. It computes the charged R\'enyi entropies \cite{BHMMMS2013}. For charged scalars in AdS$_4$ the onset on fully backreacted charged hyperbolic black holes was computed numerically in \cite{BHMM2014}.

\paragraph{Metric at first order.} We set $L=R=1$, write $r=1/u$ and take the metric as
\begin{equation}
ds^2=-F\,e^{-\Phi}dt^2+\frac{dr^2}{F}+r^2dH_3^2,
\end{equation}
which is eq.~\eqref{eq:metric} for $\Phi=0$ and $F=f/u^2$. The Einstein equations are $G_{ab}-6g_{ab}=\ve\,T_{ab}$, which fixes the normalization of the matter stress tensor. On the hairy saddle \eqref{eq:family} the $tt$ and $rr$ components of the Einstein equations give $\Phi=\frac{2\ve}{3}(eL\rhob)^2+O(\ve^2)$ at every $\mu$ and $T$, and the remaining component is a first-order equation for the correction of $F$, sourced by the probe energy density, whose solution is rational in $r$. The energy density of section~\ref{sec:b} is $3/2\kappa_5^2$ times the first-order coefficient of $-1/r^2$ in $-g_{tt}$, which at $\mu=0$ exceeds that in $F$ by $\frac23\ve p^2$, as the tracelessness of the boundary stress tensor requires. All components of the Einstein equations at order $\ve$ were verified by substitution, for symbolic $\mu$ and $\uh$. At $\mu\neq0$ the normal phase is eq.~\eqref{eq:RN}; we checked symbolically that it solves the Einstein--Maxwell equations exactly and that its temperature is $T_0/n$.

\paragraph{Grand potential at first order.} Write the bulk action as $S=S_g+\ve S_m$, with the probe solution $(g_0,\phi_0)$ for the metric and the matter fields and $g=g_0+\ve g_1$. Since $\phi_0$ extremizes $S_m$ on $g_0$, the matter corrections drop out of the terms of order $\ve^2$ in the on-shell action, and the equation $\delta^2S_g[g_1,\cdot\,]=-\delta_gS_m[\,\cdot\,]$ for $g_1$ makes them $\frac12\delta_gS_m[g_1]$, the first correction to the probe term $\ve S_m$. Per unit volume of $H^3$ the grand potential is then $\Om=\Om_g+(L/e^2)(\Om_m+\ve\,\Om_2)+O(\ve^2L/e^2)$, with $\Om_g$ that of the uncharged hyperbolic black hole, $\Om_m$ the probe result and
\begin{equation}
\Om_2=-\frac14\Big[\int_{r_h}^\infty dr\,r^3\,\bar T^{ab}\,\tilde g_{1\,ab}-2r_h^3\,\bar T{}^r{}_r(r_h)\,\delta r_h\Big],
\label{eq:Om2int}
\end{equation}
where $\bar T$ is the probe stress tensor, $\tilde g_1$ the first-order metric at fixed $(T,\mu)$, $r_h=1/\uh$ and $\delta r_h$ the shift of the horizon that keeps $T$ fixed, whose moving end gives the second term. Its coefficient is fixed independently on the Reissner--Nordstr\"om black hole \eqref{eq:RN}, where it is the only contribution, $\Om_2=-2\mu^4/[3(2+v)]$. To keep the integrand finite at the horizon we write it as $(\bar T^t{}_t-\bar T^r{}_r)F_1/F_0-\bar T^t{}_t\Phi_1$, with $F_1$ the first-order correction of $F$ at fixed $T$ and $\Phi_1$ that of $\Phi$. On $H^3$ the result is rational in $(v,\mu^2)$ plus one term proportional to $\ln(1+\mu^2\uh^2/4)$. It equals the Reissner--Nordstr\"om value at the onset and reduces to eq.~\eqref{eq:renyi1} at $\mu=0$; the full expression is long and is available from the author on request. At $\mu=0$ the ratio $W$ has its minimum $-1.80969$ at $v=2(\sqrt3-1)$, $n=2.258$; it is $-1.79676$ at $n_*$, $-1.78509$ at $n=20$ and $-\frac{16}{9}-\frac{\sqrt2}{9n}+O(n^{-2})$ at large $n$. The first-order factor of $\Delta h_n$ in eq.~\eqref{eq:dh1} is smallest, $1-1.813\,\ve$, at $n=1.755$; it is $1-1.812\,\ve$ at $n=2$, $1-1.785\,\ve$ at the merger and $1-\frac{16}{9}\ve$ in both limits.

\paragraph{Scalar-type perturbations.} For the operators of $\Dsb$ all fields are perturbed with one scalar harmonic, $-\nabla^2_{H^3}Y=\eta\,Y$: the metric, the amplitude $\delta\rho\,Y$ and the gauge-invariant vector $(b_t,b_r)\,Y$, $b_s\,\partial_iY$. The 11 linear Einstein, Maxwell and scalar equations were derived symbolically for a general slicing, with the background equations used to eliminate derivatives of the background. They are free of leftover angular dependence, and a pure-gauge perturbation with its induced matter perturbations solves all of them identically; both checks are numerical at random rational data. We use the Regge--Wheeler gauge, $\delta g_{ti}=\delta g_{ri}=0$ with $\delta g_{ij}=2r^2H_L\,\gamma_{ij}Y$ free of traceless parts, so that the remaining components $(h_{tt},h_{tr},h_{rr},H_L)\,Y$ are the gauge-invariant variables of \cite{KodamaIshibashi2003}; here $\gamma_{ij}$ is the metric of the unit $H^3$. The traceless angular equation gives $h_{rr}=e^\Phi h_{tt}/F^2-2H_L/F$, as in vacuum, since the matter has no traceless stress at linear order. For $k=0$ the time-odd fields $(h_{tr},b_r,b_s)$ decouple, and the time-even ones reduce to six first-order equations. For $k=1$ we keep the radial Hamiltonian constraint as a constraint instead of solving it for $h_{tt}$, whose coefficient can vanish inside the solution once $\ve\neq0$; with the other first-order equations imposed and $\omega\neq0$, the Bianchi identities imply the remaining equations and preserve the constraint. At the horizon the regular solutions behave as $(r-r_h)^{k/2}$, shifted by $0$ or $-1$; for $k=1$ the direction that violates the constraint is regular too and is projected out. Modes have no source for the order parameter, the phase or the external field strength, and leave the boundary metric unchanged. The eigenvalues $\eta$ are roots of the determinant of the boundary sources of the regular solutions.

\paragraph{Homogeneous onset.} Near the onset $\Om_\SF-\Om_\N=-(L/8e^2)(\mu^2-\muc^2)^2$, with $\muc^2$ shifted at order $\ve$, so the shift follows from the slope of $\Om_2^\SF-\Om_2^\N$ at the probe onset. With the closed form of $\Om_2$ this gives
\begin{equation}
\muc^2R^2=\frac{1+\sqrt{1+8n^2}-4n^2}{2n^2}+\ve\,\sigma(v),\qquad
\sigma(v)=-\frac{32\big[3v^3-8v^2+v+4-2(4-v^2)\ln(2-v)\big]}{3v\,(1-v)^2(2+v)},
\label{eq:muc1}
\end{equation}
derived for $n<1$, where $v<1$ and $\muc^2>0$. Two limits check it. As $n\to1$, $\sigma\to0$, as required by section~\ref{sec:brmu0}. As $v\to0$, the planar limit, $\muc^2$ is multiplied by $1+\frac{16}{3}(2\ln2-1)\ve$; this is the first-order shift of the critical temperature of the planar superfluid, $\mu/\pi\Tc=2+\frac{16}{3}(2\ln2-1)\ve$, found in \cite{BBM2011}, once the $1/e^2$ there is identified with $2\ve$. Backreaction raises $\muc$ at fixed $n<1$.

\paragraph{Normalizable onset.} The mode with $\eta=1$ sets in at $\mu_{\rm norm}(n)$, with $\mu_{\rm norm}\to0$ at $n_*$ \cite{MainPaper}. It has no closed form; we computed it from the static mode on the charged black hole at finite $\ve$ (table~\ref{tab:onsets}). Backreaction raises $\mu_{\rm norm}$ at the four indices computed, and the shift decreases toward $n_*$, where it must vanish by section~\ref{sec:brmu0}.

\begin{table}[t]
\centering
\small
\begin{tabular}{@{}c c c c@{}}
\toprule
$n$ & $\mu_{\rm norm}R$ ($\ve=0$) & $\partial_\ve(\mu_{\rm norm}^2R^2)$ & ratio \\
\midrule
$0.5$ & $2.330018$ & $7.67207$ & $1.413$ \\
$1$ & $1.386799$ & $1.90797$ & $0.992$ \\
$3$ & $0.604964$ & $0.178858$ & $0.489$ \\
$5$ & $0.313168$ & $0.030810$ & $0.314$ \\
\bottomrule
\end{tabular}
\caption{Onset of the normalizable instability in the probe limit and its first-order shift; the last column is the ratio $\partial_\ve\mu_{\rm norm}^2/\mu_{\rm norm}^2$.}
\label{tab:onsets}
\end{table}

\paragraph{Flux.} The flux curve \eqref{eq:alphac} is the continuation $\muc^2=-a^2$, with $\alpha=2\pi naR$. Since $\Om_2$ depends on $\mu$ only through $\mu^2$, eq.~\eqref{eq:muc1} continues to $1<v<2$, and
\begin{equation}
\alpha_c^2=\pi^2\Big[16n^2-\big(1+\sqrt{1+8n^2}\big)^2\Big]-4\pi^2n^2\,\ve\,\sigma(v)=\frac{32\pi^2}{3}\,(n-1)\Big(1+\frac{8\ve}{9}\Big)+O\big((n-1)^2\big).
\label{eq:alphac1}
\end{equation}
Backreaction raises $\alpha_c$ for $n<4.2114$ and lowers it beyond. The curve matters only while $\alpha_c<\pi$ (section~\ref{sec:flux}), that is, for $n<1.08813-0.0691\,\ve$ at first order.

\section{Numerical methods}
\label{app:numerics}

The programs for the numerical computations and the symbolic checks were written by Claude Opus 5.5 (Anthropic) at the author's direction. Units are $R=L=e=1$ and $P=eL\rho$.

\paragraph{Defect spectrum.} For a mode $\phi=u^2(1-u^2/\uh^2)^{k/2}\chi(u)\,e^{ik\tau/n}Y$, the radial equation divided by $u^3(1-u^2/\uh^2)^{k/2}$ has coefficients regular on $[0,\uh]$; the pole of the Matsubara term at the horizon cancels identically for the frequency $k/n$, which we checked symbolically. Chebyshev collocation turns it into a generalized eigenvalue problem $A\chi=\eta\,u\chi$. The weight vanishes at $u=0$, where the row is a regularity condition, and that row is removed by a Schur complement. The eigenvalues converge to 12 digits, and at $n=1$ they are the exact $\eta=0,\,-8,\,-24$ for $k=0$, the dimensions $2,4,6$ of $O$ and its transverse descendants, and $\eta=-3$ for $k=1$. The root of $\eta=1$ is $n_*=6.584444374$. Table~\ref{tab:spectrum} lists the lowest primaries; for $8\le n\le30$ the amplitude operator of $\Dsb$ with $k=0$ has $\Dh-2=0.561$--$0.569$. With flux, the background is $\Bb_\tau=a(1-u^2/\uh^2)$, and the line $n_*(\alpha)$ of section~\ref{sec:flux} is the root of $\eta=1$ in $n$ at fixed $\alpha$, for $k=0$, to $10^{-10}$; the same operator gives $\eta=0$ on the curve \eqref{eq:alphac} to $10^{-13}$. For $k=1$ the largest $\eta$ on the line is at most $0.58$, well below the merger at $\eta=1$.

\begin{table}[htb]
\centering
\small
\begin{tabular}{@{}lccccc@{}}
\toprule
$n$ & $\Dz$, $k=0$ & $\Dz$, $k=1$ & $\Dz$, $k=2$ & $\Dsb$ ampl., $k=0$ & $\Dsb$ ampl., $k=1$\\
\midrule
1 & 2 & 3 & 4 & 2 & 3\\
1.5 & 1.765236 & 2.547576 & 3.253227 & 2.330498 & 2.795962\\
2 & 1.629647 & 2.308843 & 2.869500 & 2.435389 & 2.710746\\
3 & 1.460934 & 2.048172 & 2.466479 & 2.509533 & 2.640673\\
4 & 1.344493 & 1.900144 & 2.248962 & 2.535631 & 2.612730\\
6 & 1.137604 & 1.725365 & 2.007050 & 2.554428 & 2.590617\\
8 & $1\pm0.190\,i$ & 1.617151 & 1.867426 & 2.561065 & 2.582116\\
12 & $1\pm0.313\,i$ & 1.476412 & 1.700279 & 2.565846 & 2.575589\\
\bottomrule
\end{tabular}
\caption{Lowest defect primaries at $\mu=0$ with transverse spin $k/n$, from \eqref{eq:dhat}. For $\Dz$ these are charged operators of the R\'enyi defect; for $\Dsb$ the amplitude operators. The tilt of $\Dsb$, not listed, has $\Dh=2$ exactly.}
\label{tab:spectrum}
\end{table}

\paragraph{Flows.} The collocation method uses Chebyshev collocation in $u$ and $s$, with the operator $\mathcal{L}_u$ of the spectrum computation at $k=0$. The finite-difference method derives the operator separately, in the variable $x=u^2/\uh^2$ with $P=\uh^2x\,\chi$ and divided by $\uh^2x^2$, which leaves the rows at $x=0$ and $x=1$ regular and nonzero; it uses fourth-order differences, one-sided at the ends, and starts from tanh walls rather than from the collocation solutions. Both solve by Newton's method, and the homogeneous saddle solves both discretizations to $10^{-12}$. Table~\ref{tab:walls} lists the walls of the lowest branch. The two methods agree on them to $4\times10^{-3}$, and the walls do not move when $S$ is changed from $40$ to $70$.

\begin{table}[htb]
\centering
\small
\setlength{\tabcolsep}{4.5pt}\begin{tabular}{@{}lcccccccccccc@{}}
\toprule
$n$ & 7.5 & 8 & 8.5 & 9 & 10 & 11 & 12 & 14 & 17 & 20 & 24 & 30\\
\midrule
$\nu$ & 0.157 & 0.190 & 0.215 & 0.236 & 0.268 & 0.293 & 0.313 & 0.343 & 0.375 & 0.398 & 0.420 & 0.443\\
$s_w^*$ & 20.55 & 17.07 & 15.11 & 13.83 & 12.23 & 11.25 & 10.58 & 9.72 & 8.98 & 8.54 & 8.18 & 7.84\\
$s_w^*-\pi/\nu$ & 0.505 & 0.504 & 0.505 & 0.507 & 0.516 & 0.528 & 0.541 & 0.568 & 0.608 & 0.645 & 0.690 & 0.746\\
\bottomrule
\end{tabular}
\caption{Source-free flows: $\nu$ from the spectrum, with $\Dh=1\pm i\nu$, and the wall position $s_w^*=\ln(z_*/\delta)$ of the lowest branch, where $\chi(\uh/2,s)=p/2$. The walls are converged.}
\label{tab:walls}
\end{table}

\paragraph{Branches.} Both methods find the second branch, seeded by shifting the lowest one outward by $\pi/\nu$ and adding a small negative lobe near the cutoff, and both land on it after a single step from $n=12$ to $20$, so the continuation in $n$ uses steps of at most $0.5$. The separate computation gives the offset $\pi/\nu$ between the two branches to $4\times10^{-4}$, and a direct solution in radial modes gives the ratio of their actions as $e^{-2\pi/\nu}$ to $4\times10^{-6}$.

\paragraph{Pinning.} At $n=4$, followed from $\chi_{\rm UV}=p$ toward smaller pinning, the rate $ds_w/d\ln(p/\chi_{\rm UV})$ tends to $1.5255$, which is $1/(2-\Dh_+)$ to $10^{-4}$, with $\Dh_+$ from table~\ref{tab:spectrum}.

\paragraph{Linear region and fits.} Projected onto the nodeless radial mode, the profile between the cutoff and $s_w^*-6$ is $e^s\sin(\nu s)$, with $\nu$ from the spectrum to about $10^{-4}$ and a cosine admixture below $3\times10^{-4}$. For the $\langle O\rangle$ wall the separate computation finds $c_0=0.769243$, from a fit to its walls and, independently, from the solution at $n_*$ itself, which agree to $2\times10^{-6}$; our wall lies farther in by $0.251$ at $n_*$, so $c_0=0.518$ for it. At $n=7.5$, eq.~\eqref{eq:zstar} underestimates $z_*/\delta$ by a factor of about three.

\paragraph{Ball model.} The collocation method uses $\partial_\zeta^2+2\coth\zeta\,\partial_\zeta$, a Neumann row at $\zeta=0$ and $\chi=0$ at $\zeta_{\max}$, seeded by the planar flow in the distance $\zeta_{\max}-\zeta$ from the cutoff, at $n=8$, $12$, $20$ and $30$; the finite-difference method uses fourth-order differences in $x$ and $\zeta$, at $n=12$. The wall equals the planar $s_w^*$ to $2\times10^{-3}$, and in units of $\pi\Om_{\rm hom}/2$ the area term is $1.4\times10^{-7}$ at $n=20$ and $6.5\times10^{-7}$ at $n=30$. The quadrature of the integrals $I_k$ reproduces $\Om_{\rm hom}=-\frac18\muc^4$ to $10^{-12}$.

\paragraph{Robin cutoff.} For sections~\ref{sec:bkt} and~\ref{sec:sphere} the condition at the cutoff is also replaced by $\partial_s\chi=\Gamma\chi$ on the plane and $-\partial_\zeta\chi=\Gamma\chi$ on the ball, at $\Gamma=3$ and $2$ (only $\Gamma>1$ is used); $\Gamma\to\infty$ is $\chi=0$. The position of the wall is a soft mode, so each run starts from the solution with $\chi=0$ at the same $n$, moved toward the cutoff by $\arctan\big(\nu/(\Gamma-1)\big)/\nu$, which satisfies the condition to leading order. On the plane the computed shifts agree with this prediction to $10^{-3}$ in both methods, and with $\chi=0$ the same runs reproduce the earlier walls to $5\times10^{-7}$. On the ball at $\Gamma=2$, at $n=20$ and $30$, the slope of $-2\pi I_0$ in $\zeta_{\max}$ falls tenfold for each increase of $\zeta_{\max}$ by $4$, and the area term grows to $9.4\times10^{-7}$ and $4.3\times10^{-6}$ of the homogeneous one, respectively.

\paragraph{Stability.} The eigenfunctions of $J$ with Dirichlet data in $s$ grow like $e^s$, a dynamic range of $e^S$ on the grid, so the collocation method diagonalizes $J$ conjugated by the diagonal matrix of $e^s$, which has the same eigenvalues, and the finite-difference method uses shift-and-invert on the sparse matrix. On homogeneous backgrounds, where separation of variables gives the largest eigenvalue $\eta-1-(\pi/S)^2$, both reproduce it to $7\times10^{-3}$. On the flows, at $S=40$, the largest eigenvalues are $-0.0047$, $-0.0215$, $-0.0439$ and $-0.0601$ for the lowest branch at $n=8$, $12$, $20$ and $30$, respectively. The real-time modes of section~\ref{sec:sourcefree} come from their own computation, by Chebyshev collocation in $u$ and fourth-order differences in $s$ in ingoing Eddington--Finkelstein variables, with the flows reached by continuation in a pinning field; on the homogeneous saddles its modes agree with those of a purely radial computation to $1.4\times10^{-6}$.

\paragraph{Separate computation.} This computation was written without access to the implementations of the collocation and finite-difference methods, and its solvers were fixed before its results were compared with theirs. It solves the problem on the ball model of $H^3$ in the geodesic radius, with $\chi=0$ at the cutoff and only regularity at the center. Solver I expands in the eigenfunctions of the radial operator of the normal phase and uses adaptive collocation in the radius, with the exact linear solution in the layer next to the cutoff for small $\nu$; Solver II minimizes a finite-difference discretization of the action, by gradient flow and then Newton's method. It reproduces the bifurcation of the nodeless branch from $\chi=0$ at $\zeta_{\max}=\pi/\nu$ to $3\times10^{-5}$. It finds the nodeless source-free solution at all 23 indices tested between $n=6.59$ and $30$, with the $\langle O\rangle$ wall at $s_w=244.43$ at $n=6.59$, and its walls agree with those of the finite-difference method to $2\times10^{-3}$. They reach $\nu=0.013$ and give the coefficient of $1/\nu$ equal to $\pi$ to $1.5\times10^{-10}$.

\paragraph{Flows with flux.} The flows of section~\ref{sec:flux} solve the scalar equation and the Maxwell equation for $\Bb_\tau(u,s)$ together, in the gauge in which the scalar is real, with $\Bb_\tau=a$ at the AdS boundary, $\Bb_\tau=0$ at the horizon, and the normal phase with flux, $\chi=0$ and $\Bb_\tau=a(1-u^2/\uh^2)$, at the cutoff. One method uses Chebyshev collocation in $u$ and fourth-order differences in $s$; the other, written separately, uses fourth-order differences in $x=u^2/\uh^2$ and in $s$. The winding saddle solves both discretizations to $10^{-12}$, and the two methods agree on the walls to $4\times10^{-5}$. The walls give the coefficient of $1/\nu$ equal to $\pi$ to $2\times10^{-6}$. The stability operator couples the amplitude and $\Bb_\tau$. On the ball the coefficient of $\ln(2R/\delta)$ agrees with that of the winding saddle to $3\times10^{-8}$. Below the line, at $\alpha=\pi$ and $n=5$ and $6$, Newton's method does not converge from any of four initial walls, and at $n=6.68$ the flow at $\alpha=0$, followed in $\alpha$, runs off as $\pi/\nu$ as $\alpha$ approaches the line. On the flows of section~\ref{sec:flux}, where $n\ge7.5$ and so $aR\le1/2n\le0.067$, the screening term, integrated over the flow with the weight $e^{-2s}$ of the action, is at most $0.22\%$ of the quartic repulsion of the potential.

\paragraph{Real chemical potential.} The flows at real chemical potential solve eqs.~\eqref{eq:flowmu} by Chebyshev collocation in $u$, fourth-order differences in $s$ and Newton's method, with $B/B_{\rm N}$, which is regular at the horizon, as the unknown. On the plane the far end is held at $\Dsb$ at $s=S$; at $(n,\mu)=(2,2)$ and $(1,3)$ the walls are converged to $4\times10^{-6}$ and do not change, to eight digits, between $S=20$ and $40$. Fitting $s_w$ with the terms $1/\nu$, $1$ and $\nu^2$ to walls with $\nu$ down to $0.05$ gives the coefficient of $1/\nu$ equal to $\pi$ to $1.0\times10^{-4}$, relative, at $n=1$ and $n=3$. With the pinning field $\chi_{\rm UV}$ at the cutoff, the wall moves at $ds_w/d\ln(\bar\chi/\chi_{\rm UV})=3.33011$ at $(n,\mu)=(1,1)$ and $1.44533$ at $(4,0.2)$, the values of $1/(2-\Dh_+)$ there to the digits shown. On the ball at $(n,\mu)=(2,1.2)$ the amplitude at the center is $0.22\,\bar\chi$ for $\zeta_{\max}=1.02\,\pi/\nu$ and $10^{-24}\bar\chi$ for $0.98\,\pi/\nu$, and at $(4,1)$, $(2,2)$, $(1,2)$ and $(1,3)$ the coefficient of $\ln(2R/\delta)$ equals $-2\pi\Om_{\rm hom}$ to $10^{-8}$, relative. The reduced action is convex at the midpoints of $90$ pairs of random profiles, and with the sign of imaginary $\mu$ it fails on $21$ of $30$. The tail of the flow at $(1,3)$ decays at $\Dh-2=2.33346-0.43628\,i$, against $2.33347-0.43626\,i$ for the leading pair of $\Dsb$. The line $\mu_{\rm norm}(n)$ is the root of $\eta=1$ for the radial operator of the normal phase at $\mu$, discretized as for the defect spectrum.

\paragraph{Backreaction.} The nonlinear hairy black holes are shot from the horizon with a Runge--Kutta integrator in $\ln r$, with the data at the horizon solved for a source-free scalar and for the targets $(\mu,T)$, at five values of $\ve$; the derivatives at $\ve=0$ come from polynomial fits. At $\mu=0$ their first-order entropy agrees with eq.~\eqref{eq:renyi1} to $6\times10^{-10}$ at five indices between $n=1.2$ and $20$. For the onsets the scalar is linear on the charged black hole \eqref{eq:RN}; the perturbation equations are integrated in $\ln r$, with the variables scaled by their falloffs at infinity, and the eigenvalues $\eta$ (appendix~\ref{app:backreaction}) are found by secant iteration. At $n=\frac14$, $\frac12$ and $0.8$ the zero mode gives $\partial_\ve(\muc^2R^2)$ within $1.1\times10^{-6}$, relative, of eq.~\eqref{eq:muc1}, and the onsets of table~\ref{tab:onsets} change by at most $2\times10^{-7}$, relative, between the cutoffs $u=10^{-4}$ and $10^{-5}$. For the amplitude operators of $\Dsb$ the derivatives in $\ve$ use five values of $\ve$, and at $\ve=0$ the integration reproduces every amplitude dimension of table~\ref{tab:spectrum} to the six digits given there.

\paragraph{Genericity.} For the quartic coupling of section~\ref{sec:lambda}, the London equation of section~\ref{sec:tilt} on the saddle is solved by Chebyshev collocation in $u^2$, which at $\lambda=e^2/2$ reproduces the rational solution \eqref{eq:London} to $10^{-12}$. At $n=2$ it gives $C_{\tilt}/C_{\tilt}|_{\lambda=e^2/2}=2.54$, $1.63$, $0.577$, $0.316$ and $0.085$ for $\lambda/e^2=\frac18$, $\frac14$, $1$, $2$ and $8$, respectively. For a scalar of dimension $\Delta$ the field equation is that of section~\ref{sec:crossing}. The spectrum, by Chebyshev collocation in $x$, gives $n_*=1.467$, $2.228$, $3.616$, $6.584$, $14.82$ and $52.65$ at $\Delta=1.25$, $1.5$, $1.75$, $2$, $2.25$ and $2.5$, respectively, to $5\times10^{-6}$ in relative terms, and the dimensions at which $n_*=2$, $3$ and $4$ to $10^{-8}$. At the onset of the hairy phase the ratio of the slopes of $\Dh_{\rm amp}$ and of the charged operator of $\Dz$ is $-2$ to $2\times10^{-4}$ at every $\Delta$ computed. For $\Delta<2$ the free energy of the source-free homogeneous saddle, per unit volume of $H^3$ and in units of $L/e^2$, is $\Om_\SF-\Om_\N=-\frac{c_3}{4}\int_0^{\uh}du\,u^{4\Delta-5}\chi^4$. The usual counterterm, quadratic in $P$ on the cutoff surface, cancels the only divergence, and the Legendre transform to the alternate quantization adds a term proportional to the product of the source and the one-point function, which vanishes without a source. Shooting and Chebyshev collocation in $\ln u$ give it to $10^{-13}$, and at $\Delta=2$ they reproduce the closed form.

The flows use a mixed method, Chebyshev collocation in $x$ and fourth-order differences in $s$, solved by damped Newton iteration. The first flow at each $\Delta$ comes from continuation in a pinning field, $\chi(x,0)=t\chi_{\rm hom}(x)$ with $t$ from $1$ to $0$, the others from continuation in $n$. At $\Delta=2.25$ the finite-difference method, extended to general $\Delta$, reproduces the walls to $2\times10^{-4}$ up to $n=35$. Fits of $s_w^*$ with the terms $1/\nu$, $1$, $\nu^2$ and $\nu^4$ over $\nu\le0.2$ give the coefficient of $1/\nu$ equal to $\pi$ to $3.3\times10^{-6}$. For the $\langle O\rangle$ wall, the solution at $n_*$ itself, in radial modes with the exact linear layer, gives $c_0=0.109$, $1.292$ and $2.177$ at $\Delta=1.5$, $2.25$ and $2.5$, respectively, against $0.769$ at $\Delta=2$; the fits agree with these to $6\times10^{-4}$. At $\Delta=1.25$ the walls at $n=2$ and $3$ move by up to $1.4\times10^{-2}$ under refinement, so only the existence of these flows is used. On the ball, with nothing imposed in the interior, the wall agrees with the planar one to $2\times10^{-5}$. Stability uses the largest eigenvalues of the linearized operator, with the controls used at $\Delta=2$.

\paragraph{Bound on the quartic.} Near the AdS boundary the black hole and the gradients along $H^3$ drop out, and after the counterterm quadratic in $P$ the action per unit volume of $H^3$ is $\int du\,u^{-5}\big[\frac12(u\partial_uP-\Delta P)^2+\frac14c_3P^4+\frac16c_5P^6\big]$. Completing the square turns it into
\begin{equation*}
\int du\,u^{-5}\Big[\frac12\Big(u\partial_uP-\Delta P-\frac{c_3P^3}{4(\Delta-1)}\Big)^2+\Big(\frac{c_5}{6}-\frac{c_3^2}{32(\Delta-1)^2}\Big)P^6\Big]
\end{equation*}
in pure AdS, with no term from the AdS boundary for $\Delta>1$. This is nonnegative for $c_3^*\le c_3<0$, with $c_3^*=-(\Delta-1)\sqrt{16c_5/3}$, and for $c_3\ge0$ the action is nonnegative term by term. At $c_3^*$ the wall on which the square vanishes has zero action. Below $c_3^*$ the same wall has the action $\frac14(c_3-c_3^*)\int du\,u^{-5}P^4<0$, and moving it toward the AdS boundary multiplies that by any large factor. For $c_5=\frac1{10}$, $c_3^*=-0.730$ at $\Delta=2$ and $-0.913$ at $\Delta=2.25$.

\paragraph{Attractive quartic.} In this paragraph and the next three, $s_w$ is the $\langle O\rangle$ wall of section~\ref{sec:cutoff}. For the attractive quartic ($c_3=-\frac12$, $c_5=\frac1{10}$) at $\Delta=2$ the walls beyond $n_*$ lie far below $\pi/\nu$, with $s_w=19.42$ at $\nu=0.0793$, where $\pi/\nu=39.6$. Fitted, with a term in $\nu^2$, to the seven walls with $\nu<0.2$, at $n=6.6$ to $8$, the matching law gives $\ell=50.1$ and $c=2.71$, and it predicts a flow at $n_*$ with $s_w=52.80$ and flows below $n_*$ down to $n_c=6.5712$. The mixed method and a Galerkin method, described below, give $s_w=52.80$ at $n_*$ and flows below $n_*$ at least down to $n=6.574$, where $s_w=82.1$, with walls within a relative $1.3\times10^{-4}$ of the law's.

Below $n_*$ the mixed method reaches the flows by continuation in the pinning field and Newton's method. A solution counts only if it is nodeless and positive, its residual is below $10^{-7}$ and its wall lies at least $15$ inside the domain. Closer to $n_c$ the walls exceed $95$, and on longer domains these runs do not converge, so they do not bracket $n_c$. The Galerkin method expands $\chi$ in a Jacobi basis in $x$ and solves the resulting system in $s$ by adaptive collocation; it reproduces the walls of the mixed method to $1.5\times10^{-5}$, relative.

At $\Delta=2.25$, with $c_3=-0.7$ and $c_5=\frac1{10}$, both methods find the flows at $n=14$, $13$ and $12$, with walls that agree to $8\times10^{-4}$, and the mixed method also finds one at $n=11.96875$ but none at $n=11.96094$, beyond the end of the branch.

A further computation locates that end. In radial modes, with the exact decay at the homogeneous saddle and the exact linear solution next to the cutoff, the flow is the single orbit from the saddle into the linear region, so no domain limits the wall. There the nodeless mode is $e^s(c_+e^{\bar\nu s}+c_-e^{-\bar\nu s})$, which has a zero, where the cutoff can sit, only for $c_-/c_+<0$. The branch therefore ends at $c_-=0$. This gives $n_c=6.57116$ at $\Delta=2$ and $n_c=11.9628$ at $\Delta=2.25$, to $2\times10^{-5}$. Just above $n_c$ the wall diverges as $-\ln(n-n_c)/(2\bar\nu)$, the power law of section~\ref{sec:pinu}: as $n-n_c$ falls from $10^{-3}$ to $3\times10^{-5}$, $2\bar\nu s_w+\ln(n-n_c)$, with $\bar\nu$ at $n_c$, changes by only $2\times10^{-3}$ at $\Delta=2.25$, and by $0.11$ at $\Delta=2$, where $\bar\nu$ itself changes by $4\%$ over that range.

\bibliographystyle{unsrturl-compact}
\bibliography{refs}

\end{document}